\documentclass{2023SCGE}

\usepackage{graphicx}	
\usepackage{amsmath}	
\usepackage[table,xcdraw,dvipsnames]{xcolor}
\usepackage{hyperref}
\usepackage{float}
\usepackage{multirow}

\DeclareGraphicsExtensions{.pdf,.png,.jpg}

\DeclareRobustCommand{\ion}[2]{%
\relax\ifmmode
\ifx\testbx\f@series
{\mathbf{#1\,\mathsc{#2}}}\else
{\mathrm{#1\,\mathsc{#2}}}\fi
\else\textup{#1\,{\mdseries\textsc{#2}}}%
\fi}

\begin{document}

\ensubject{subject}

\ArticleType{Article}
\SpecialTopic{SPECIAL TOPIC: }
\Year{2023}
\Month{January}
\Vol{66}
\No{1}
\DOI{??}
\ArtNo{000000}
\ReceiveDate{January 11, 2023}
\AcceptDate{April 6, 2023}

\title{MUltiplexed Survey Telescope (MUST) Science White Paper I: Overview of Large-Scale Structure Cosmology in the Era of Stage-V Spectroscopic Surveys}

\author[1]{\\Cheng Zhao\thanks{Email: \href{mailto:czhao@tsinghua.edu.cn}{czhao@tsinghua.edu.cn}}}{}%
\author[1]{Song Huang\thanks{Email: \href{mailto:shuang@tsinghua.edu.cn}{shuang@tsinghua.edu.cn}}}{}
\author[1,2]{Mengfan He}{}
\author[3]{Paulo Montero-Camacho}{}
\author[1]{Yu Liu}{}
\author[1]{Pablo Renard}{}
\author[1]{\\Yunyi Tang}{}
\author[4]{Aur\'{e}lien Verdier}{}
\author[1]{Wenshuo Xu}{}
\author[1]{Xiaorui Yang}{}
\author[4]{Jiaxi Yu}{}
\author[1]{Yao Zhang}{}
\author[1]{Siyi Zhao}{}
\author[5,6]{\\Xingchen Zhou}{}
\author[4]{Sheng-Yu He}{}
\author[4]{Jean-Paul Kneib}{}
\author[1,7]{Jiayi Li}{}
\author[1]{Zhuoyang Li}{}
\author[8,9]{Wen-Ting Wang}{}
\author[10]{\\Zhong-Zhi Xianyu}{}
\author[1,7]{Yidian Zhang}{}
\author[4]{Rafaela Gsponer}{}
\author[11]{Xiao-Dong Li}{}
\author[4]{Antoine Rocher}{}
\author[12]{\\Siwei Zou}{}
\author[13]{Ting Tan}{}
\author[11]{Zhiqi Huang}{}
\author[1]{Zhuoxiao Wang}{}
\author[1]{Pei Li}{}
\author[4]{Maxime Rombach}{}
\author[3]{\\Chenxing Dong}{}
\author[4]{Daniel Forero-S\'anchez}{}
\author[14,1]{Yuanhang Ning}{}
\author[2]{Huanyuan Shan}{}
\author[15,16]{Tao Wang}{}
\author[3]{\\Yin Li}{}
\author[8,9]{Zhongxu Zhai}{}
\author[5,17]{Yuting Wang}{}
\author[5,18,17,12]{Gong-Bo Zhao}{}
\author[15,16]{Yong Shi}{}
\author[1]{\\Shude Mao}{}
\author[1]{Lei Huang}{}
\author[1]{Liquan Guo}{}
\author[1]{Zheng Cai}{}

\AuthorMark{Zhao}

\AuthorCitation{Zhao, et al}

\address[1]{Department of Astronomy, Tsinghua University, Beijing 100084, China}
\address[2]{Shanghai Astronomical Observatory, Chinese Academy of Sciences, 80 Nandan Road, Shanghai 200030, China}
\address[3]{Pengcheng Laboratory, Nanshan District, Shenzhen, Guangdong 518000, China}
\address[4]{Institute of Physics, Laboratory of Astrophysics, \'{E}cole Polytechnique F\'{e}d\'{e}rale de Lausanne (EPFL), Observatoire de Sauverny,\\ CH-1290 Versoix, Switzerland}
\address[5]{National Astronomical Observatories, Chinese Academy of Sciences, Beijing 100101, China}
\address[6]{Science Center for China Space Station Telescope, National Astronomical Observatories, Chinese Academy of Science,\\ 20A Datun Road, Beijing 100101, China}
\address[7]{Zhili College, Tsinghua University, Beijing 100084, China}
\address[8]{Department of Astronomy, Shanghai Jiao Tong University, Shanghai 200240, China}
\address[9]{Shanghai Key Laboratory for Particle Physics and Cosmology, Shanghai 200240, China}
\address[10]{\; Department of Physics, Tsinghua University, Beijing 100084, China}
\address[11]{\; School of Physics and Astronomy, Sun Yat-sen University, Zhuhai 519082, China}
\address[12]{\; Chinese Academy of Sciences South America Center for Astronomy (CASSACA), National Astronomical Observatories of China,\\ Beijing 100101, China}
\address[13]{\;IRFU, CEA, Universit\'{e} Paris-Saclay, F-91191 Gif-sur-Yvette, France}
\address[14]{\;Department of Scientific Research, Beijing Planetarium, Beijing 100044, China}
\address[15]{\;School of Astronomy and Space Science, Nanjing University, 163 Xianlin Avenue, Nanjing 210023, China}
\address[16]{\;Key Laboratory of Modern Astronomy and Astrophysics, Nanjing University, 163 Xianlin Avenue, Nanjing 210023, China}
\address[17]{\;Institute for Frontiers in Astronomy and Astrophysics, Beijing Normal University, Beijing 102206, China}
\address[18]{\;School of Astronomy and Space Science, University of Chinese Academy of Sciences, Beijing 100049, China}

\abstract{
    The MUltiplexed Survey Telescope (MUST) is a 6.5-meter telescope under development. Dedicated to highly-multiplexed, wide-field spectroscopic surveys, MUST observes over 20,000 targets simultaneously using 6.2-mm pitch positioning robots within a $\sim 5\,{\rm deg}^2$ field of view. MUST aims to conduct the first Stage-V spectroscopic survey in the 2030s, mapping the 3D Universe with over 100 million galaxies and quasars, spanning from the nearby Universe to a redshift of $z\sim 5.5$, corresponding to approximately 1 billion years after the Big Bang. To cover this extensive redshift range, we present an initial conceptual target-selection algorithm for different types of galaxies, including local bright galaxies, luminous red galaxies, emission-line galaxies, and high-redshift ($2 < z < 5.5$) Lyman-break galaxies. Using Fisher forecasts, we demonstrate that MUST can address fundamental questions in cosmology, including the nature of dark energy, tests of gravity theories, and investigations into primordial physics. This is the first paper in a series of science white papers for MUST, with subsequent papers focusing on additional scientific cases, including galaxy and quasar evolution, Milky Way physics, and dynamic phenomena in the time-domain Universe.
}

\vspace{1cm}

\vspace{0.2cm}

\maketitle

\tableofcontents

\begin{multicols}{2}

\section{Introduction}
\label{sec:intro}

Over the past four decades, beginning with the ``Stick Man'' from the CfA Redshift Survey in 1982 \cite{Huchra83, Lapparent86}, spectroscopic mapping of large-scale structures (LSS) has accumulated more than 30 million redshifts of nearby and distant galaxies. This monumental achievement has contributed significantly to the establishment of the current cosmological model $\Lambda$CDM, along with other cosmological probes, i.e., cosmic microwave background (CMB; \cite{Bennett13, Planck2020c}), Type-Ia supernovae (SN; \cite{Pantheon+2022, Betoule2014, DESY5_2024}), or measurements of weak lensing (e.g., \cite{DES2022, Kids1000_2021}). For two decades (2000--2020), major experiments such as the Sloan Digital Sky Survey (SDSS;~\cite{eBOSS2021}), followed by the ongoing (2021--2026) survey from the Dark Energy Spectroscopic Instrument\footnote{\url{https://www.desi.lbl.gov/the-desi-survey/}} (DESI;~\cite{DESI2022overview}), have mapped the 3D universe at low and intermediate redshift ($ z\lesssim 3$). Clustering measurements from spectroscopic surveys of galaxies and quasars have become a key probe of cosmology. They provide precise measurements on the baryon acoustic oscillations (BAO) scale~\cite{Eisenstein2005} and the linear growth rate of structure $f\sigma_8$ through redshift space distortions (RSD) \cite{RSD_Kaiser_1987}. Recent results from the DESI collaboration \cite{DESI2024a} suggest a potential deviation from the cosmological constant to time-varying dark energy. By the end of this decade, we expect to have sub-percent-level constraints on dark energy and gravity from galaxy clustering out to redshift $z\sim2$. When the DESI project concludes, we will have completed the spectroscopic survey component of the four stages outlined in the Dark Energy Task Force (DETF) report \cite{Albrecht2006DETF}. Yet, many fundamental questions remain unanswered, calling for a new era of cosmological experiments in the 2030s. 

Going one step further, 3D maps of the universe at high redshift ($z > 2$) will enable the observation of linear modes in the primordial universe, significantly enhancing our ability to constrain dark energy and inflation \cite{Sailer+2021}. In the next decade, a series of ground- and space-based photometric surveys CSST \cite{Zhan2021}, Euclid \cite{Euclidcoll2024}, Nancy Grace Roman Space Telescope \cite{WFIRST2013}, LSST \cite{LSST} will provide deep and high-quality images for future galaxy spectroscopic surveys allowing to target galaxies such as Lyman Break Galaxy (LBG) or Lyman-$\alpha$ emitter (LAE) \cite{Ruhlmann-Kleider2024, Payerne2024} as tracer of matter at high redshift~$2<z<5$. Large-volume, high-redshift surveys using these new tracers have the unprecedented potential to test primordial non-Gaussianity, probe dynamic dark energy, reveal new features in the primordial power spectrum, and uncover tantalizing hints of new physics. 

At the same time, a high-density spectroscopic survey of low-redshift ($z<1.5$) can provide a high-fidelity 3D map of the cosmic web and trace matter distributions into the non-linear regime, opening up unexplored scientific opportunities. Such a multi-purpose dataset can also enhance the scientific performance of other cosmological probes, such as calibrating the photometric redshift and intrinsic alignment models for weak gravitational lensing surveys (e.g., \cite{McCullough2024}; \cite{Samuroff2023, Lamman2023}), providing spectroscopic follow-up and host galaxy properties for supernova surveys (e.g., \cite{Wolf2016, Soumagnac2024}), or exploring new approaches to map the low-redshift large-scale structures (LSS) such as a peculiar velocities survey (e.g., \cite{Koda2014, Said2024}). More importantly, it will help maximize the potential for synergies between spectroscopic surveys and other cosmological probes, such as CMB (e.g., \cite{2021MNRAS.501.6181K}), weak lensing (e.g., \cite{Myles2021, Yuan2024, Lange2024, Chen2024}), and intensity mapping (IM, e.g., \cite{Amiri2023}) experiments. 

Motivated by these two promising directions, the cosmological \& high-energy physics community has recently coined the concept for a Stage-V spectroscopic experiment (e.g., \cite{Chang2022, Annis2022}) to fulfill these high expectations. By definition, a Stage-V spectroscopic facility should use a telescope with high etendue ($A\Omega$; $A$ and $\Omega$ denote the collecting area and field of view of the telescope, respectively) to achieve high survey speed. More importantly, the facility should have significantly improved multiplexed capability (i.e., the number of targets that can be observed simultaneously) compared to the Stage-IV survey (e.g., 5,000 fibers for DESI) \cite{Cai2025}. Conceptually, a Stage-V facility requires a minimum of 10,000 fibers that can be readily reconfigured to target different objects. This is the primary technical challenge now for such an ambitious vision. At the same time, a Stage-V facility should also have excellent optical performance, high-performance multi-object spectrographs that cover at least the entire optical wavelength range, and a site with favorable observing conditions. Building on these requirements, multiple ground-based concepts have been proposed, including the 6.5\,m MegaMapper telescope \cite{Schlegel22MegaMapper, Blanc22}, the dual-6\,m \& dual-hemisphere Spec-S5 project\footnote{\url{https://www.spec-s5.org/}} \cite{P5}, the 11~m Maunakea Spectroscopic Explorer\footnote{\url{https://mse.cfht.hawaii.edu/}} (MSE; \cite{MSE}), the 12~m Wide-field Spectroscopic Telescope\footnote{\url{https://www.wstelescope.com/}} (WST; \cite{Mainieri2024}), and the $\sim$12~m Extremely Large Spectroscopic Survey Telescope (ESST; \cite{Su2024}). 

The MUltiplexed Survey Telescope\footnote{\url{https://must.astro.tsinghua.edu.cn/en}} (MUST) is a 6.5-meter telescope~\cite{Zhang23} under active development. MUST will be located at the 4358~m Peak A of Saishiteng Mountain in Qinghai, China. Equipped with over 20,000 fibers over a $\sim5~\rm{deg}^2$ field of view (FoV), it features three-channel spectrographs covering wavelengths from 370-960 nm, with spectral resolution between $R = 2,000$ and $4,500$. MUST is designed to conduct an ambitious Stage-V cosmological spectroscopic survey, aiming to precisely measure key cosmological parameters and improve our understanding of dark energy and cosmic evolution. With the first light scheduled for 2031, MUST expects to conduct the first Stage-V spectroscopic survey, targeting Lyman Break Galaxy (LBG) and Lyman-$\alpha$ emitter (LAE) at high redshift across $\sim 13,000$~deg$^2$ of the northern sky. Clustering analysis of these tracers will provide sub-percent precision measurements on BAO parameters -- $D_{\rm A}(z)/r_{\rm d}$ and $H(z)r_{\rm d}$ -- and the linear growth rate of structure, $f\sigma_8$, at redshift $2<z<5$, a region not yet covered by current spectroscopic galaxy surveys. Additionally, MUST will provide stringent constraints on primordial non-Gaussianities (PNG, local type) through the parameter $f_{\mathrm{NL}}^{\mathrm{local}}$  with a precision of $\sigma(f_{\mathrm{NL}}^{\mathrm{local}}) \sim 1$. This will enable stringent testing of a wide range of inflationary models. MUST is expected to provide sufficient precision ($\sim 0.03\,{\rm eV}$ when combined with CMB data) on the sum of neutrino masses to constrain a nonzero neutrino mass with $\sim 2\,\sigma$ significance, assuming normal hierarchy. The inverted mass hierarchy can be tested with a significance $\sim 1.3\,\sigma$. Finally, from power spectrum measurements of the Lyman-$\alpha$ forest from quasars, MUST will yield the most precise constraint on warm dark matter mass to date $m_X > 10.5$\,keV at 95\% confidence level (assuming 14,000~deg$^2$ and $k_{\rm max} = 0.67$ $h$\,Mpc$^{-1}$). 

Besides the unique potential of the MUST project, synergies with other cosmological surveys, i.e., future imaging surveys (e.g., CSST, Euclid, LSST), CMB experiments (e.g., Simon Observatory \cite{SO2018}, CMB-S4 \cite{CMB-S4}, LiteBird \cite{LiteBIRD}) or radio surveys (e.g., SKAO \cite{SKA2020}) will enhance the constraining power of MUST and will result in a better understanding of our Universe. As a dedicated spectroscopic survey facility, MUST can also conduct surveys that support a wide range of scientific topics beyond LSS cosmology, including galaxy evolution, supermassive black holes (SMBHs), the structure of the Milky Way, and time-domain astrophysics. 

This paper describes the MUST instrument and the scientific objectives of the cosmological survey that will be conducted over a 5-year period of observation. Section~\ref{sec:must} provides an overview of the MUST project, including the current status of the whole project, the design of the telescope (Section~\ref{ssec:telescope}), the focal plane system \& the spectrograph (Section~\ref{ssec:instrument}), the site \& observing condition (Section~\ref{ssec:site}), and the overall scientific capabilities (Section~\ref{ssec:scidrive}). Section~\ref{sec:motivation} describes the key scientific motivations of MUST for the Stage-V cosmological surveys. The primary scientific goals are covered in detail, with a brief summary of the potential for new probes and synergies with other cosmological surveys. Section~\ref{sec:target} presents the current target selection strategy and provides the redshift distribution and target density estimations for the cosmological forecast. We will also introduce the conceptual survey design for MUST. Finally, Section~\ref{sec:forecast} describes the method for theoretically forecasting the cosmological potential of MUST and summarizes the forecast for dark energy, structure growth, primordial non-Gaussianity, neutrino mass, and warm dark matter constraints. Discussions and main conclusions of this project and future directions are described in Section~\ref{sec:conclusion}.

Throughout this work, we adopt as fiducial baseline $\Lambda$CDM cosmology with parameters~$H_0=67.6$~km~s$^{-1}$ Mpc$^{-1}$, ~$\Omega_b=0.046$ and $\Omega_m=0.31$. All magnitudes in this work are defined in the AB magnitude system \cite{Oke83}.

\section{MUltiplexed Survey Telescope}
    \label{sec:must}

\begin{figure*}[!htb]
    \centering
    \includegraphics[width=1.0\linewidth]{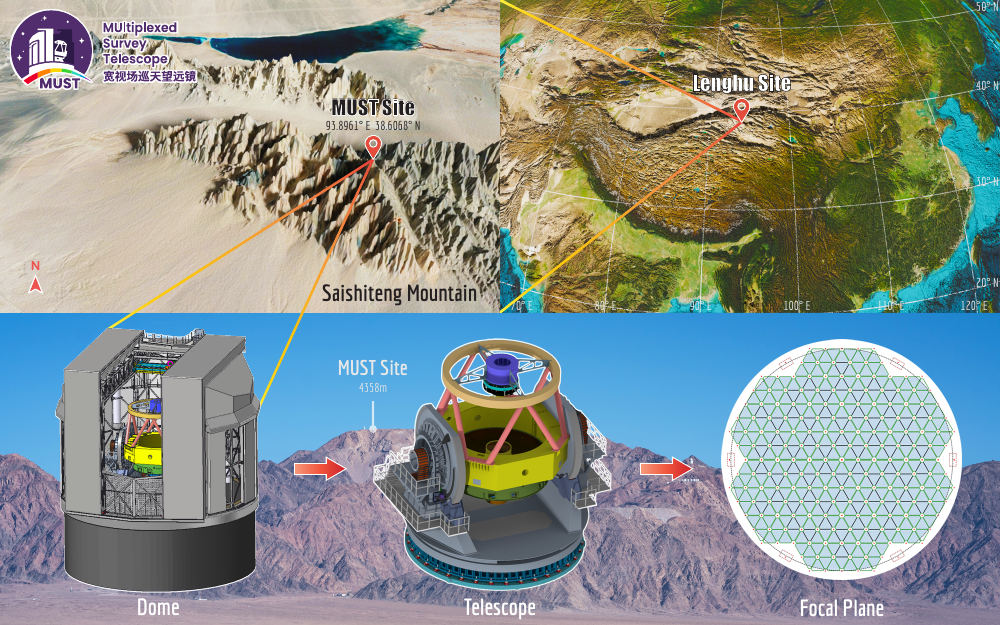}
    \caption{
        Overview of the MUST project. The top-right panel shows the location of the currently selected MUST site in Qinghai Province, China. The top-left panel and the background image of the bottom panel feature photographs of Saishiteng Mountain near Lenghu. The highest peak, at 4358 m, was selected as the site for MUST. 
        From left to right, the bottom panels illustrate the preliminary design of the dome of MUST and the telescope, as well as a sketch of the conceptual design of the focal plane of MUST. Under the current design, MUST will host 21,168 robotic fiber positioners, each using 336 triangular modules. 
    }
    \label{fig1}
\end{figure*}

\begin{table*}[!htb]
\centering
\resizebox{12.5cm}{!}{%
\begin{tabular}{|
>{\columncolor[HTML]{EFEFEF}}c ccc|}
\hline
\multicolumn{4}{|c|}{\cellcolor[HTML]{DAE8FC}Optical System} \\ \hline
\multicolumn{1}{|c|}{\cellcolor[HTML]{EFEFEF}Optical Design} &
  \multicolumn{3}{c|}{R-C with WFC} \\ \hline
\multicolumn{1}{|c|}{\cellcolor[HTML]{EFEFEF}Primary Mirror Diameter} &
  \multicolumn{1}{c|}{6.5 m Hyperboloid} &
  \multicolumn{2}{c|}{With 2 m central hole} \\ \hline
\multicolumn{1}{|c|}{\cellcolor[HTML]{EFEFEF}Secondary Mirror Diameter} &
  \multicolumn{1}{c|}{2.4 m Hyperboloid} &
  \multicolumn{2}{c|}{Convex} \\ \hline
\multicolumn{1}{|c|}{\cellcolor[HTML]{EFEFEF}Wide Field Corrector} &
  \multicolumn{1}{c|}{Five-Lens Design} &
  \multicolumn{2}{c|}{Largest lens diameter: 1.6 m} \\ \hline
\multicolumn{1}{|c|}{\cellcolor[HTML]{EFEFEF}Throughput Requirement} &
  \multicolumn{1}{c|}{$>50$\% from 370-960 nm} &
  \multicolumn{2}{c|}{\begin{tabular}[c]{@{}c@{}}Considering the reflectance of the primary \& secondary mirror, \\ WFC throughput, and the vignetting of the secondary \& corrector\end{tabular}} \\ \hline
\multicolumn{1}{|c|}{\cellcolor[HTML]{EFEFEF}Focal Ratio} &
  \multicolumn{1}{c|}{F/3.7} &
  \multicolumn{2}{c|}{Cassegrain Focus} \\ \hline
\multicolumn{1}{|c|}{\cellcolor[HTML]{EFEFEF}Field of View} &
  \multicolumn{1}{c|}{2.8$^{\circ}$ in Diameter} &
  \multicolumn{2}{c|}{} \\ \hline
\multicolumn{4}{|c|}{\cellcolor[HTML]{CBCEFB}Focal Plane \& Fiber System (Preliminary)} \\ \hline
\multicolumn{1}{|c|}{\cellcolor[HTML]{EFEFEF}Focal Plane Diameter} &
  \multicolumn{1}{c|}{1.2 m} &
  \multicolumn{2}{c|}{} \\ \hline
\multicolumn{1}{|c|}{\cellcolor[HTML]{EFEFEF}Total \# of Modules} &
  \multicolumn{1}{c|}{336} &
  \multicolumn{2}{c|}{Semi-frameless design} \\ \hline
\multicolumn{1}{|c|}{\cellcolor[HTML]{EFEFEF}\# of Positioners per Module} &
  \multicolumn{1}{c|}{63} &
  \multicolumn{2}{c|}{21 $\times$ 3 groups} \\ \hline
\multicolumn{1}{|c|}{\cellcolor[HTML]{EFEFEF}Total \# of Fiber Positioners} &
  \multicolumn{1}{c|}{21,168} &
  \multicolumn{2}{c|}{} \\ \hline
\multicolumn{1}{|c|}{\cellcolor[HTML]{EFEFEF}Effective Positioner Coverage} &
  \multicolumn{1}{c|}{74.0\%} &
  \multicolumn{2}{c|}{} \\ \hline
\multicolumn{1}{|c|}{\cellcolor[HTML]{EFEFEF}Pitch Distance between Positioners} &
  \multicolumn{1}{c|}{6.2 mm} &
  \multicolumn{2}{c|}{Using a $\theta$-$\phi$ design} \\ \hline
\multicolumn{1}{|c|}{\cellcolor[HTML]{EFEFEF}Fiber Core Diameter} &
  \multicolumn{1}{c|}{140 $\mu$m} &
  \multicolumn{2}{c|}{1.2 arcsec on the sky} \\ \hline
\multicolumn{1}{|c|}{\cellcolor[HTML]{EFEFEF}Fiber Route Length} &
  \multicolumn{1}{c|}{45-50 m} &
  \multicolumn{2}{c|}{From the fiber tip to the slithead} \\ \hline
\multicolumn{4}{|c|}{\cellcolor[HTML]{FFFFC7}Spectrograph (Preliminary)} \\ \hline
\multicolumn{1}{|c|}{\cellcolor[HTML]{EFEFEF}Channel} &
  \multicolumn{1}{c|}{\cellcolor[HTML]{EFEFEF}Wavelength Coverage} &
  \multicolumn{1}{c|}{\cellcolor[HTML]{EFEFEF}Spectral Resolution} &
  \cellcolor[HTML]{EFEFEF}Average Throughput \\ \hline
\multicolumn{1}{|c|}{\cellcolor[HTML]{96FFFB}Blue} &
  \multicolumn{1}{c|}{370-590 nm} &
  \multicolumn{1}{c|}{R$\sim 1900$-3300} &
  $\geq 55$\% \\ \hline
\multicolumn{1}{|c|}{\cellcolor[HTML]{FFCE93}Red} &
  \multicolumn{1}{c|}{565-775 nm} &
  \multicolumn{1}{c|}{R$\sim 3300$-4500} &
  $\geq 60$\% \\ \hline
\multicolumn{1}{|c|}{\cellcolor[HTML]{FFCCC9}NIR} &
  \multicolumn{1}{c|}{750-960 nm} &
  \multicolumn{1}{c|}{R$\sim 4300$-5500} &
  $\geq 60$\% \\ \hline
\end{tabular}%
}
\caption{
    Summary of the key specifications of the optical, focal plane \& fiber, and spectrograph systems of MUST. 
    Please note that the design specifications for the focal plane and spectrograph systems are still preliminary. 
    }
    \label{table1}
\end{table*}

The MUltiplexed Survey Telescope (MUST) is a dedicated spectroscopic survey facility proposed and led by the Department of Astronomy at Tsinghua University and co-founded with \'{E}cole Polytechnique F\'{e}d\'{e}rale de Lausanne (EPFL). 

The MUST project aims to build a 6.5-meter-wide-field telescope with multiplexed spectroscopic capabilities by 2030. While designed as a flexible platform for various spectroscopic surveys, the primary scientific drive of MUST is to become the first Stage-V spectroscopic survey \cite{Schlegel22stage5} that answers fundamental questions in cosmology and physics. The concept and early design of MUST were inspired by the MegaMapper project \cite{Schlegel22MegaMapper, Blanc22}. The MUST collaboration has independently completed the preliminary design of the optical system \cite{Zhang23}, structure, and dome \cite{Marchiori24} of the telescope, and is working with collaborators and vendors to design the modular focal plane, fiber, and spectrograph systems \cite{Li24}. Here, we provide a brief introduction to the design and technical capabilities of MUST. A summary of the key specifications is available in Table~\ref{table1}. An upcoming paper will summarize a more detailed description and analysis of the MUST project's technical aspects.

\subsection{A 6.5-m Telescope for Spectroscopic Surveys}
    \label{ssec:telescope}

Driven by the demanding scientific requirements of next-generation cosmological surveys, a Stage-V spectroscopic telescope requires excellent optical quality over a large Field of View (FoV) on a telescope with a diameter greater than 6\,m. The optical design of MUST is driven by the need to address these technical challenges. Currently, MUST adopts a compact Ritchey-Chretien (R-C) design with a multi-element Wide Field Corrector (WFC). The hyperboloid primary and secondary mirrors of MUST are 6.5\,m and 2.4\,m in diameter. The primary mirror has a 2\, m-diameter central hole to facilitate installation of a five-lens WFC, ensuring excellent imaging quality for MUST. The largest lens for the WFC of MUST is 1.6\,m in diameter, even slightly larger than the largest lens of the camera of LSST \cite{LSST}. MUST published the conceptual optical design in 2023 \cite{Zhang23}. Since then, the collaboration has implemented a series of modifications, primarily to the WFC configuration, to improve the engineering feasibility of the telescope and reduce risk during the manufacture and assembly of critical subsystems (Zhang et al., in preparation). Under the updated design, MUST imaging quality over the entire 2.8$^{\circ}$ diameter FoV is excellent: up to a 50$^{\circ}$ zenith angle and within the wavelength range of 0.365 to 1.0\,$\mu$m, the 80\% Encircled Energy (EE80) size of the image spot is $<0.6$\,arcsec. At the Cassegrain focus, which hosts MUST's modular focal plane system, the optical system has a focal ratio of F/3.7. 

\subsection{Multiplexed Focal Plane \& Instruments}
    \label{ssec:instrument}

Enabled by an optimized optical design, MUST can achieve significantly improved multiplexing capability compared with current Stage-IV spectroscopic facilities, using a novel modular focal-plane design (See Figure~\ref{fig1}). Starting from the LAMOST survey, modern spectroscopic surveys have adopted different types of robotic fiber positioners (\cite{Xing98, Smith04, Akiyama08, Staszak16, Brzeski18}) to guide the light from distant targets to the scientific instruments and enable efficient \& flexible survey design. The largest multiplexed survey facility, DESI, currently hosts 5,000 fiber positioners, each with an outer diameter of 10\,mm \cite{Schubnell16}. Each positioner was installed and operated independently on the DESI focal plate. However, as a Stage-V facility, MUST has a much more demanding requirement for multiplexed capability; individually managing more than 10,000 fiber positioners significantly increases the complexity and risk of instrument operation. 

MUST is collaborating with EPFL and industrial partners to develop a novel modular focal plane system \cite{Rombach24}. This modular design was originally proposed for the MegaMapper project by \cite{Silber2022} and is also being developed for the Spec-S5 project \cite{SpecS5}. The MUST design greatly benefits from the open-source design and the model file\footnote{The model files are available at \url{https://zenodo.org/records/6354859}}, released by \cite{Silber2022}. The Trillium design is one of the candidates that is being tested for MUST.

For MUST, as shown in the bottom-right panel of Figure~\ref{fig1}, a 1.2-m-diameter focal plate will host 336 triangular fiber-positioner modules at the Cassegrain focus under the current preliminary design. Each module will integrate 60 or 63 fiber-positioning robots, grouped according to whether three positioning references are required for each module. MUST plans to adopt a miniaturized fiber-positioning robot with an outer diameter of 6.2\,mm and an optical fiber with a $\sim$140\,$\mu$m core diameter, corresponding to 1.2 arcsecs on the sky, to ensure that fiber density meets the requirements of future cosmology surveys. To optimize fiber coverage, the focal plate groups four modules together. The gaps between modules are 1 and 3\,mm within the group and between adjacent groups. Altogether, this allows MUST to equip 20,160 or 21,168 fiber positioners over the focal plane with a 74\% coverage, which results in a $\sim 5$\,deg$^2$ FoV covered by fibers with a $\sim$4,000\,deg$^{-2}$ fiber density. It is worth noting that, given the aspheric shape of the focal plane and the requirements of the fiber throughput \& focal ratio degradation (FRD) budgets of MUST, the fiber tips of all the positioners should be located within $\pm 100\,\mu{\rm m}$ from the theoretical focal plane. The design of MUST achieves this goal by approximating the focal plane with a best-fit spherical surface that meets the requirement. In this way, all modules can have identical configurations, and within each module, the tips of the 60- or 63-positioners will match the spherical surface. This challenging and ambitious design is essential to MUST's overall scientific capability, particularly for the LSS survey discussed here. 

The first-generation instruments of MUST consist of $\sim42$ multi-object spectrographs; each will host $\sim500$ optical fibers on a 4k$\times$4k-pixel CCD detector. The preliminary design of MUST's spectrograph is still underway. Regardless, given the very similar scientific goals to the DESI project, the current concept adopts a three-channel design similar to the DESI spectrograph, covering the wavelength range from $\sim3700$ to $\sim9600$\,\AA{} with a spectroscopic resolution of $R\sim1,900$--5,500. We currently aim to achieve an average throughput of $>60$\% across all three channels while exploring different approaches to further improve it to an average of $\sim 70$\%. 

In addition to the fiber positioning system and the spectrographs, the scientific instrument system of MUST includes other crucial sub-systems, such as the fiber view camera (FVC), fiducial fibers, focal plate adjustment \& derotation mechanism, and a complex fiber route (see Figure~\ref{fig1}). Altogether, they will enable MUST to realize the potential of its Stage-V LSS survey. In forthcoming publications, we will describe the technical design of the entire focal plane system and the spectrographs.

\subsection{Site and Observing Conditions}
    \label{ssec:site}

MUST has selected Peak A of Saishiteng Mountain, located near Lenghu Town in Haixi Mongol and Tibetan Autonomous Prefecture, Qinghai Province, China, as its candidate site (referred to as the ``Lenghu'' site). 
The Lenghu site was first reported in \cite{Deng2021} and has been selected by a series of domestic astronomical projects in China, including the Mozi Survey Telescope\footnote{\url{https://wfst.ustc.edu.cn/}} (a 2.5\,m wide field imaging survey telescope in operation) and the Jiao-tong University Spectroscopic Telescope (JUST; a planned 4\,m segmented mirror telescope; \cite{JUSTTeam2024}), and more. Along with several other peaks (see Figure~\ref{fig1}), Peak A, at 4,358\,m, has been flattened for construction. Starting in Oct. 2023, the MUST collaboration began monitoring the weather and observing conditions on Peak A. While we are still accumulating data for a more precise site condition assessment, combining our data with the public data collected by the NAOC team at the nearby Peak C\footnote{\url{http://lenghu.china-vo.org/}}, we estimate that the Lenghu site has an annual clear night fraction between 60\% to 70\%, with worse observing conditions during the summer, and a median DIMM seeing FWHM between 0.8 to 1.0\,arcsec. These specifications meet the site requirements for conducting a fiber-spectroscopic survey. While light pollution from nearby towns and mining businesses is a concern, the local government has passed legislation to protect dark night conditions within a 50\,km radius of the Lenghu site. 

At 4,358\,m, Peak A enjoys significantly less atmospheric attenuation in the shorter wavelength range. Preliminary photometric observations from the Mozi telescope indicate a promising result: the atmospheric attenuation at $\lambda < 450$\,nm for Lenghu is comparable to that at Mauna Kea and significantly better than at lower-altitude sites, such as Kitt Peak. This presents MUST an opportunity to improve the overall throughput in the blue wavelength end, which is crucial for identifying the Lyman-$\alpha$ (Ly$\alpha$) emission line at $z>2$. Meanwhile, the high altitude also results in a median nighttime air temperature of -8.1$^{\circ}$C (minimum \& maximum temperatures are -25$^{\circ}$C and 15.2$^{\circ}$C), which increases the challenges for the construction and maintenance of MUST. At the same time, the median nighttime humidity is $\sim30$\% at Peak A. 

\subsection{Survey Capability and Overall Scientific Potential}
    \label{ssec:scidrive}

\begin{figure*}[!htb]
\centering
    \includegraphics[width=1.0\linewidth]{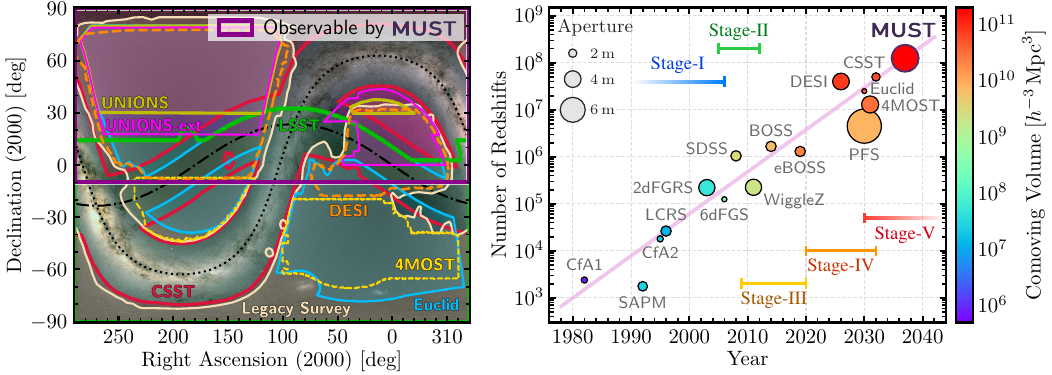}
    \caption{
        {\it Left} panel: full-sky map showing the region observable by MUST (declination above $-10^\circ$), overlaid with the footprints of relevant imaging surveys (Legacy Survey, CSST, UNIONS, LSST, and Euclid) and spectroscopic surveys (DESI and 4MOST) on top of the stellar density map from {\it Gaia} EDR3 \cite{Gaia_EDR3}. The black dotted and dash-dotted lines show the Galactic and ecliptic planes, respectively.
        {\it Right} panel: accumulation of galaxy and quasar redshifts from spectroscopic surveys since the early 1980s. The aperture of the telescopes scales the sizes of circles used, and filled colors represent the comoving volumes probed by direct LSS tracers (excluding Ly$\alpha$ forest tracers). Also shown are the rough divisions between stages of the spectroscopic surveys. MUST aims to become the first Stage-V survey in operation by the early 2030s.
    }
    \label{fig:surveys}
\end{figure*}

Given the current monthly statistics of clear night fraction, the Lenghu site could provide $ \sim $ 2,400 and 2,800 observing hours per year. Assuming a configuration with 20,000 working fibers and a $>90$\% uptime, MUST will have 210-270 million fiber hours in a five-year survey, allowing MUST to conduct Stage-V LSS surveys (see Figure~\ref{fig:surveys}). As a dedicated survey facility, MUST's long-term commitment to spectroscopic surveys is another unique strength in fulfilling the Stage-V cosmological goals. Compared to DESI, MUST has a 3.4 times higher light collecting capability and 4$\times$ more fibers. Even considering the throughput costs of the secondary mirror, the more complex WFC design, the longer fiber route, and challenges in the spectrograph design, MUST can still improve the spectroscopic survey efficiency -- the number of redshifts measured for the same set of targets during the same time -- by an order of magnitude. 

While the Stage-V cosmological survey will be the highest priority of MUST during the first phase of its operation, the significant fiber-hour budget will allow us to design versatile programs with a wide range of scientific goals. In addition to mapping the large-scale structures for cosmology, the same dataset will enable robust statistical studies of galaxies and Active Galactic Nuclei (AGN) near and far. These data can help us understand the rise and fall of star formation in galaxies over the last 10 Billion Years and the assembly of different galaxy populations.  It will also enable us to measure the accretion rate and mass of supermassive black holes (SMBHs) in a galaxy sample that is one order of magnitude larger than the current one, significantly improving our understanding of SMBH growth and their impact on galaxy evolution. At the small, non-linear scale, a detailed picture of galaxy clustering will enable better modeling of galaxy-halo connections, which in turn will benefit cosmology. During bright nights, MUST can measure the radial velocities and chemical abundances of many halo stars in the Milky Way, shedding light on its assembly history and constraining the nature of dark matter. Located at a longitude with fewer large ground-based telescopes, MUST is also poised to play an exciting role in the age of time-domain spectroscopic surveys. The list can go on. 

In addition, the modular focal plane of MUST secures future opportunities for instrument upgrades, as a high-resolution fiber spectrograph and an integrated field spectrograph (IFS) can be straightforwardly added to the focal plane, provided their front-end optics fit into the current fiber positioner module. In the following publications from this series, we will discuss the scientific potential and strategy of these topics in detail. 

For the remainder of this work, we will focus on the primary scientific goal of MUST: the Stage-V LSS spectroscopic survey, which aims to deepen our understanding of several fundamental questions in cosmology and physics. 

\section{Scientific Motivations of the Stage-V Cosmological Surveys}
    \label{sec:motivation}

Ever since the discovery of the accelerated expansion of the Universe using Type Ia supernova data \cite{Riess1998, Perlmutter1999}, $\Lambda$CDM has been widely accepted as the standard cosmological model (cf., however, \cite{Carroll1992} for earlier observational evidence of a nonzero $\Lambda$). However, several key components of the $\Lambda$CDM model remain unknown, including the physical origin of $\Lambda$ (or, more generally, dark energy), the nature of dark matter, and the establishment of initial conditions for cosmic structures. Meanwhile, several observational challenges to $\Lambda$CDM have emerged as the precision of cosmological measurements improves (e.g., \cite{Perivolaropoulos2022}). Massive spectroscopic surveys are expected to help address these issues. By analyzing the clustering of large-scale structures (LSS) from spectroscopic data, we can investigate the properties of dark energy and dark matter through the dynamic evolution of the Universe governed by these competing components and extract signatures of fundamental physical processes at an extremely high energy scale in the primordial Universe.

\begin{figure*}[!htb]
    \centering
    \includegraphics{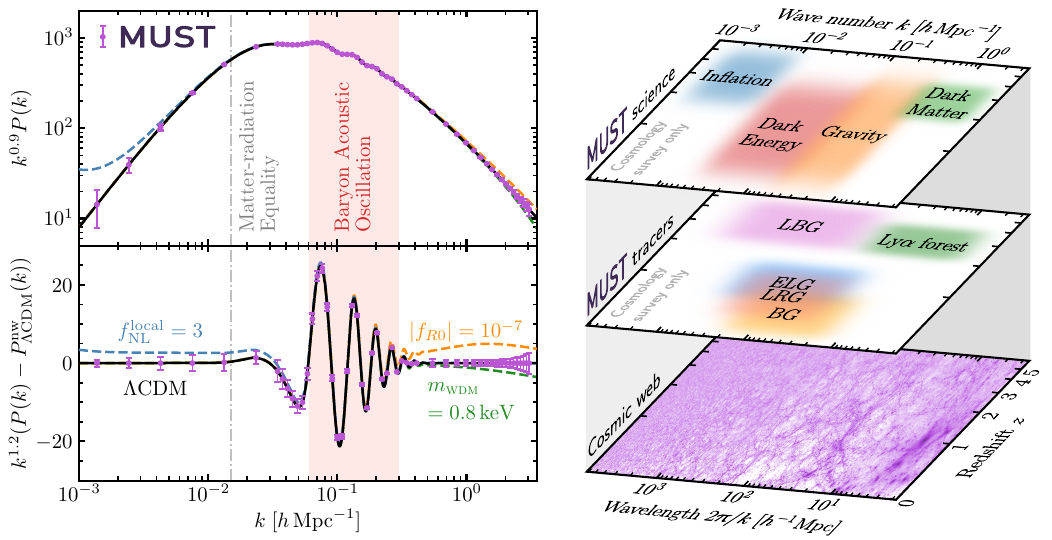}
    \caption{
        {\it Left} panel: linear galaxy power spectra for various cosmological models, with forecasted error bars based on the current survey design of MUST. The error bars are centered on the Planck 2018 $\Lambda$CDM cosmology \cite{Planck2020l}, and the models include additional components that reflect the key science cases of MUST: local-type primordial non-Gaussianity with $f_{\rm NL}^{\rm local} = 3$, $f(R)$ gravity with $|f_{\rm R 0}| = 10^{-7}$, and warm dark matter with a particle mass of $m_{\rm WDM} = 0.8\,{\rm keV}$. Error bars for $k < 0.4\,h\,{\rm Mpc^{-1}}$ are estimated using Eq.~\eqref{eq:covariance}, with the expected bias and number density of Lyman-break galaxies. For $k > 0.4\,h\,{\rm Mpc^{-1}}$, errors are from forecasts of the Ly$\alpha$ 3D power spectrum, accounting for spectroscopic systematics and the quasar luminosity function. All power spectra and error bars are rescaled to match the amplitude of the matter power spectrum in the fiducial $\Lambda$CDM model for visual clarity.
        {\it Right} panel: illustration of the key science cases ({\it top}), primary LSS tracers ({\it middle}), and the underlying dark matter density field ({\it bottom}) across different scales and redshifts targeted by MUST.
    }
    \label{fig:scicases}
\end{figure*}

In this section, we summarize the science drivers for a next-generation (Stage-V) spectroscopic survey with MUST. Figure~\ref{fig:scicases} highlights a few key scientific goals that MUST can address with high significance.
Our aim is to derive survey-level requirements from these goals so that the final design maximizes cosmological outcome within the instrument specifications and observational constraints.
A full, end-to-end optimization for every possible case is beyond the scope of this first paper, and many topics still require further development in theory and data analysis methods. We therefore concentrate on well-established cases -- dark energy and structure growth -- for detailed forecasts, and provide indicative prospects for primordial physics, the sum of neutrino masses, and the mass of warm dark matter particles in Section~\ref{sec:forecast}. Broader science topics will be developed in follow-up papers.

\subsection{Nature \& Evolution of Dark Energy}
    \label{ssec:darkenergy}

 Dark energy is one of the key scientific problems of our time, posing a significant challenge to the Standard Model of particle physics. While CMB observations suggest that dark energy constitutes about 70\,\% of the energy density of the Universe in the $\Lambda$CDM model \cite{Planck2020c}, its physical properties are largely unexplored. One exception is the equation-of-state (EoS) parameter $w$, which influences the dynamics of the Universe and can be probed through geometric measurements. The most straightforward interpretation of dark energy is the cosmological constant ($\Lambda$), corresponding to a constant $w = -1$, which is consistent with most observations to date (e.g., \cite{Planck2020c,eBOSS2021}). A natural extension beyond $\Lambda$ involves introducing dynamic dark energy, which potentially originates from various physical mechanisms \cite{Copeland2006} and leads to a time-evolving EoS parameter typically expressed as the Chevallier-Polarski-Linder (CPL) parametrization \cite{Chevallier2001, Linder2003}:

\begin{equation}
    \label{eq:de_cpl}
    w(a)=w_0+w_a(1-a),    
\end{equation}

where $w_0$ is the current value of $w$ and $w_a$ quantifies its time evolution. 
For instance, the quintessence model, which describes a dynamical scalar field minimally coupled to gravity, typically yields $w \gtrsim -1$, whereas the phantom energy, a scalar field with negative kinetic energy, predicts $w < -1$. Observational constraints on $w_0$ and $w_a$ provide a pathway to distinguishing these models.

The constraints on EoS parameters rely primarily on geometrical probes, including ``standard(-isable) candles'' such as Cepheids and type Ia supernovae (e.g., \cite{DESY5_2024}), and ``standard rulers'' such as Baryon Acoustic Oscillations (BAO; \cite{Eisenstein1998}) and the matter-radiation equality scale (e.g., \cite{Philcox2021}). BAO arises from acoustic density waves in the primordial Universe and produces an excess of matter correlation amplitude at a characteristic comoving scale of $\sim 100\,h^{-1}\,{\rm Mpc}$. By extracting the BAO signal from the clustering of matter tracers, such as galaxies and QSOs, we can probe the expansion rate of our Universe and thus constrain dark energy models. Recently, the tomographic Alcock–Paczynski (AP) test, which performs geometrical measurements based on the response of the apparent shape of galaxy clustering to the adopted cosmological model for redshift-to-distance conversion, has also been shown as a complementary probe of dark energy \cite{dong2023tomographic}.

Observational programs for measuring the dark energy EoS have been strategically planned to proceed through successive stages of cosmic surveys, as outlined in the Dark Energy Task Force report \cite{Albrecht2006DETF}. Since the first detections of the BAO signal from galaxies at $z<0.47$ \cite{Cole2005, Eisenstein2005}, subsequent spectroscopic surveys have extended BAO measurements to galaxies across a wide redshift range (up to $z<1.5$), as well as QSOs and Lyman-$\alpha$ tracers at even higher redshifts \cite{eBOSS2021}. These efforts have substantially improved the precision of $w$ measurements. Recently, DESI, a Stage-IV dark energy survey, reported a $w$ consistent with $\Lambda$CDM with an uncertainty of $\sim 2.5\%$ \cite{DESI2025}. However, when the EoS is allowed to vary over time, current spectroscopic data hint at the potential dynamical behavior of dark energy \cite{DESI2024c, DESI2025}. With its unprecedented survey efficiency, MUST is expected to deliver even more precise BAO measurements, covering a redshift range from the nearby Universe to $z\sim 5.5$. Thus, MUST meets the requirements of a Stage-V dark energy survey suggested by the Snowmass Cosmic Frontier report \cite{Chou2022} and may offer deeper insights into the potential evolution of dark energy (see Section~\ref{ssec:forecastbao}). Additionally, we will investigate the possibility of improving dark energy constraints using alternative probes, such as the tomographic AP test and the matter-radiation equality scale.

\subsection{Growth of Structure \& Nature of Gravity}
\label{ssec:gravity}

Within the standard cosmological framework, gravity amplifies primordial density perturbations of order $10^{-5}$, seeded by inflation, in an expanding cosmic background, leading to the formation and growth of large-scale structure (LSS). The structure growth history is therefore a sensitive probe of both gravity and dark energy. More generally, the growth of structure traces the dynamical response of matter to gravity, while geometrical probes constrain the cosmic expansion history. In GR these two aspects obey a consistency relation, while departures from GR can modify or break it. Tracking the growth history over cosmic time therefore provides a powerful test of gravity on cosmological scales \cite{Huterer2023}.

The growth of cosmic structure shapes both the spatial clustering of galaxies and their peculiar velocities. One key observable is redshift-space distortions (RSD; e.g., \cite{Hamilton1998}), where the observed redshifts of galaxies are perturbed by their peculiar velocities along the line of sight, producing apparent anisotropic clustering. This effect can be used to constrain the combination $f(z)\sigma_8(z)$, where $f(z) \equiv {\rm d}\ln D / {\rm d}\ln a$ is the linear growth rate, with $D(a)$ the linear growth factor, through the autocorrelation of galaxy velocity divergence $P_{\theta \theta} \propto (f \sigma_8)^2$ and its cross-correlation with galaxy densities $P_{g \theta} \propto f \sigma_8$ (e.g., \cite{RSD_revive_2009}).

Tests of GR are often formulated by comparing the measured growth history with the GR expectation, e.g., through the growth-index parameterization $f(z) \simeq \Omega_m(z)^\gamma$, for which $\gamma \simeq 0.55$ in GR-$\Lambda$CDM and can shift in alternative gravity theories (e.g., \cite{Linder2007}).
A complementary approach to testing GR is to probe potential deviations of the two scalar gravitational potentials $\Psi$ and $\Phi$ from their GR relations. Such deviations are usually parameterized by phenomenological functions $\mu$ and $\Sigma$ that linearly perturb GR (e.g., \cite{ZhaoGongBo:2010, eBOSS2021, DES:2022extension}):
\begin{align}
    k^2 \Psi &= - 4 \pi G a^2 (1 + \mu(a,k)) \rho \delta ,\label{eq:mg_mu} \\
    k^2 (\Psi + \Phi) &= -8 \pi G a^2 (1+\Sigma(a,k)) \rho \delta . \label{eq:mg_sigma}
\end{align}
Here, $\rho$ indicates the background matter density and $\delta$ is the comoving density perturbation. The function $\mu$ characterizes modifications to the motion of non-relativistic matter, while $\Sigma$ describes modifications to the lensing potential relevant for relativistic particles. In GR, $\mu=\Sigma=0$.

Current observations are broadly consistent with GR and leave only limited room for modified gravity on cosmological scales. Recent DESI-based analyses find $\mu$ and $\Sigma$ to be consistent with 0, and constraints on Horndeski-class models are likewise compatible with GR \cite{DESI2024MG}. Screening models, such as Hu--Sawicki $f(R)$ gravity \cite{fR_model} and Dvali--Gabadadze--Porrati (DGP) braneworld gravity \cite{DGP_discovery_2000} are also increasingly constrained (e.g., \cite{Vogt_fR, Vogt_nDGP}).
Moreover, GW170817 has eliminated a large fraction of the viable scalar--tensor parameter space, sharpening the case for precision tests of the surviving models \cite{Kase2019}.
Nevertheless, robust measurements of structure growth beyond $z\sim 2$ remain limited. Proving this regime is critical for distinguishing modified gravity scenarios and breaking degeneracies with dark energy models. As a Stage-V cosmological survey, MUST is expected to fill this gap and substantially improve tests of gravity.

MUST also has the potential to measure relativistic effects, such as the gravitational redshifts, that are sensitive to gravity theories but have been too subtle to be detected with current data \cite{McDonald2009}. These effects can introduce additional distortions to galaxy clustering beyond RSD, primarily generating odd Legendre multipoles in the cross-correlation of different galaxy populations that can be measured from galaxy spectroscopic data.

\subsection{Inflation \& Primordial Physics}
    \label{ssec:primordial}

Large-scale structures (LSS) of the Universe not only trace its evolution but also encode rich information about its initial conditions, formed during a primordial era. The most prominent scenario for the primordial universe is cosmic inflation, which refers to a nearly exponential expansion of the universe by roughly 60 $e$-folds in a blink. 
To probe the microscopic quantum fluctuations of spacetime and matter/fields during inflation, which led to the formation of the LSS \cite{Achucarro:2022qrl, Ellis:2023wic}, is a primary motivation for MUST.

Through setting the initial conditions of the LSS, inflationary cosmology provides a wealth of observables that reveal the dynamics of the primordial universe and, more interestingly, the fundamental physics at the inflation scale. Current theory and observations suggest that the inflation scale could be up to $\mathcal{O}(10^{13}\,\text{GeV})$ \cite{Planck:2018jri}, which is much higher than the energy scale of any imaginable terrestrial experiments. Therefore, primordial fluctuations provide a unique window into high-energy fundamental physics, complementing other high-energy experiments. 

The inflation predictions for the primordial scalar power spectrum have been delicately measured at the CMB scale through the overall amplitude, $A_s$, and the spectral index, $n_s$ \cite{Planck:2018jri}. 
These parameters are also expected to be measured by MUST by extracting the primordial scalar power spectrum from the LSS data. 
However, due to the near-scale invariance of the scalar fluctuation at large scales, the information extractable from the power spectrum is limited, although there have been ongoing efforts in searching for and understanding tiny scale-dependent (and very often oscillatory) features in the power spectrum \cite{Flauger:2009ab, Chen:2014joa}. 
On the other hand, a vast amount of information can be revealed by going to higher $n$-point ($n\geq 3$) correlations of scalar modes, which are traditionally known as primordial non-Gaussianity (PNG; \cite{Chen:2010xka, Meerburg:2019qqi}). 
There is in principle fruitful information encoded in the PNG, though historically, many efforts have been made to the search of 3-point statistics, such as bispectra $B(k_1,k_2,k_3)$ on a few particular shapes (the bispectrum dependences on the ratios of $k_i$ where $i=1,2,3$), including the local, equilateral, and orthogonal shapes (e.g., \cite{karagiannis2014png,cagliari2023png,scoccimarro2012nonlocalpng,coulton2023quijotepng}). 

The overall amplitudes of these shapes are parameterized by dimensionless parameters called $f_\text{NL}$, which are crucial cosmological parameters for the next generation LSS surveys. 
Within typical single-field slow-roll inflation models, it is worth noting that the minimal signal of equilateral bispectrum coming from the gravitational interaction of the scalar modes is $f_\text{NL}\sim \mathcal{O}(10^{-2})$, which, if detected, would mark our detection of gravitational interaction \cite{Maldacena:2002vr}. 
However, other fundamental interactions, particles, and fields besides the inflaton field may carry stronger-than-gravity coupling to the scalar mode. Thus, it is possible that more significant signals can be detected, thereby revolutionizing our understanding of the primordial universe and fundamental physics. 
Typically, $f_\text{NL}\sim 1$ is a crucial threshold for the theories with and without the slow-roll condition \cite{kawasaki2011curvaton,cai2009bounce}.

There have been active searches of $f_{\text{NL}}$, especially for the local shape ($f_{\text{NL}}^{\text{local}}$), but a firm detection has not been made yet. The state-of-the-art constraint comes from CMB measurements by Planck, with $f_\text{NL}^{\text{local}}=-0.9 \pm 5.1$ \cite{planck2020png}. Meanwhile, galaxy spectroscopic surveys are becoming increasingly promising, as LSS clustering can be highly sensitive to the PNG through scale-dependent galaxy bias \cite{Dalal:2007cu, Green:2023uyz, Achucarro:2022qrl}. 
Current bound on the local shape PNG from spectroscopic surveys is  $\left|f_{\mathrm{NL}}^{\text{local}}\right| \sim \mathcal{O}(10)$ \cite{DESI_PNG}. 
MUST, as a Stage-V spectroscopic survey, is expected to improve the constraints on $f_\text{NL}$ significantly with the benefit of the increasing survey volume and redshift range, surpassing current CMB constraints (see Section~\ref{ssec:forecastfnl}).

Besides the traditional PNG searches, recent years have witnessed rapid development in new directions, such as cosmological collider (CC) physics. 
For instance, oscillatory signals of $n$-point correlation functions can be generated by the resonance between heavy states spontaneously created during inflation and scalar (or tensor) modes and encode a wealth of information about the evolution history of spacetime and the dynamical information of the heavy particle, including the mass, spin, sound speed, chemical potential, interaction type, etc.  \cite{Chen:2009we, Arkani-Hamed:2015bza}
Meanwhile, parity-odd patterns in galaxy clustering have gained considerable attention recently. If confirmed, they will be a clear signal of new physics. 
Active efforts are ongoing to identify these signals in current data and forecasts for detectability from future observations \cite{Meerburg:2016zdz, Green:2023uyz, Cabass:2024wob, Sohn:2024xzd}.
To maximize the potential scientific outcome of MUST, we aim to explore the feasibility of achieving these goals in subsequent studies.

\subsection{Neutrinos \& Light Relics}
    \label{ssec:neutrinos}

The existence of neutrino mass, as revealed by atmospheric and solar neutrino experiments, provides strong evidence for new physics beyond the Standard Model. Neutrino oscillation experiments suggest two possible hierarchies in their mass spectrum: the normal ($m_1 < m_2 < m_3$, $M_\nu \gtrsim 0.059\,{\rm eV}$) and the inverted ($m_3 < m_1 < m_2$, $M_\nu \gtrsim 0.099\,{\rm eV}$) hierarchies, where $m_1$, $m_2$, and $m_3$ indicate the 3 mass eigenstates, and $M_\nu$ denotes the total neutrino mass \cite{Navas2024PDG}.
Determining the mass hierarchy is essential for understanding the nature of neutrinos (Dirac or Majorana) and formulating a generalized standard model. However, the most stringent constraint on $M_\nu$ from current particle physics experiments, set by the KATRIN experiment \cite{Aker2024}, is only $M_{\nu} \lesssim 1.4\,{\rm eV}$ at the 90\,\% credible level (CL). This limit is insufficient to distinguish between the mass hierarchies.

Neutrinos act as radiation in the early Universe and subsequently contribute matter-energy budget at the late Universe with $\Omega_{\nu}=M_{\nu} / (93.14\,h^{2}\,{\rm eV})$, which alters the expansion rate and shifts the redshift of matter-radiation equality. 
Besides, due to their significant velocity dispersion, massive neutrinos slow down the growth of structures, resulting in a suppression of the matter power spectrum below the neutrino free-streaming scale \cite{Lesgourgues2006, Lesgourgues2012}. As a result, LSS probes are sensitive to the total neutrino mass, providing complementary constraints to particle physics experiments.

In addition to neutrino mass, the clustering of galaxies also permits constraints on the effective number of neutrino species $N_{\rm eff}$, which can be used to probe non-standard neutrino interactions \cite{2021JHEP...05..058D, 2021PhLB..82036508D} or extra light thermal relics from the primordial Universe, such as light sterile neutrinos \cite{2012arXiv1204.5379A} and axions \cite{1977PhRvL..38.1440P, 1978PhRvL..40..223W, 1978PhRvL..40..279W}.
This is because $N_{\rm eff}$ can be inferred from the phase shift of the BAO spectrum.
Any deviation from the value $N_{\rm eff} = 3.044$, as predicted by the standard cosmological model with 3 massive neutrino species, notably if exceeding 1\%, would indicate new physics beyond the Standard Models of particle physics and cosmology.

It is worth noting that the constraints on $M_{\nu}$ and $N_{\rm eff}$ from CMB alone exhibit a geometrical degeneracy with $H_0$ (or $\Omega_m$). BAO measurements can help break this degeneracy due to their ability to constrain $\Omega_m$ \cite{Planck2020c, DESI2024c}.
Combining CMB and LSS data has already led to more stringent constraints on $M_{\nu}$ than those from terrestrial experiments, approaching the lower bound in the inverted hierarchical scenario \cite{2016JCAP...11..035H, 2018JCAP...03..011G, 2020JCAP...07..037C}.
Recently, DESI reported an upper bound of $M_{\nu} < 0.072\,{\rm eV}$ (95\% CL) with a prior of $M_{\nu} > 0\,{\rm eV}$, and updated the constraint $N_{\rm eff} = 3.10 \pm 0.17$, which is in remarkable agreement with the Standard Model expectation.
With its significantly larger survey volume and increased tracer number density, MUST is expected to substantially tighten constraints on both parameters and investigate the possibility of distinguishing between the two mass hierarchies in combination with data from next-generation CMB experiments (e.g., CMB-S4 \cite{2022arXiv220308024A}).

\subsection{Dark Matter}
    \label{ssec:darkmatter}

Although dark matter (DM) -- which constitutes $\sim 25$\,\% of the energy density of the Universe -- was first postulated almost 90 years ago \cite{Zwicky1937}, its fundamental nature remains unknown. Observational efforts have focused mainly on the cold dark matter (CDM) paradigm, yet its constituents remain elusive, prompting interest in alternative DM models. These alternative models aim to address specific challenges to the CDM model, such as the missing satellite, cusp-core (e.g., \cite{1994ApJ...427L...1F,1994Natur.370..629M}), and too-big-to-fail problems \cite{2011MNRAS.415L..40B} (see \cite{Buckley2018} for a review).

Popular alternatives to the CDM model include warm dark matter (WDM; see e.g., \cite{Schaeffer1988, Nadler2021b}), fuzzy dark matter (FDM; \cite{Hu2000, Hlozek2015}), self-interacting dark matter (SIDM; \cite{Spergel2000, Zeng2022}), and primordial black holes (PBH; see e.g., \cite{Hawking1971, Montero-Camacho2019}). These models gain traction for their potential to mitigate the challenges of CMB or for their natural origin, such as PBHs, which do not require new physics beyond the standard model. MUST can offer unique insight into the nature of dark matter through three avenues: Milky Way (MW) and Local Group (LG) observations, as well as the Ly$\alpha$ forest.

In the current {\it Gaia} \cite{Gaia2016} era, the measurement of the dark matter distribution around our Milky Way based on stellar kinematics still has significant uncertainties (e.g., \cite{2020SCPMA..6309801W}). One dominant reason is the lack of stellar tracers with full phase-space information in the outer Milky Way halo ($\gtrsim 100$~kpc). MUST can measure the line-of-sight velocities (LOSV) of a plethora of MW halo stars, which, in combination with a deep proper motion survey (e.g., the Roman space telescope high-latitude survey \cite{Lin2022}), will allow for the mapping of the local distribution of dark matter (e.g., \cite{Sanderson2015}) around our MW, enabling constraints on the radial mass profile and 3D shape of the MW dark halo, which will significantly benefit direct dark matter detection programs (e.g., \cite{Salucci2019}). 

In particular, the ongoing Stage-IV DESI Milky Way survey (MWS; \cite{2023ApJ...947...37C}) reaches $\sim20$ -- 30\,\% completeness for MW targets in the flux range of $16<r<19$. MUST can potentially reach fainter magnitudes than DESI and thus can provide more distant MW halo star LOSVs, increasing completeness at similar magnitudes to DESI and providing better constraints on the dark matter distribution in the outer halo. In addition to halo stars, dwarf galaxies and stellar streams around our MW and in the LG carry valuable information on the nature of dark matter (e.g., \cite{2012MNRAS.420.2318L,2015MNRAS.452.3650O,2017MNRAS.466..628B,2019ApJ...880...38B,2021MNRAS.502.2364B, Zeng2022,2023MNRAS.521.4630J}), and MUST can potentially measure the LOSVs for member stars in distant dwarf galaxies or streams to better infer their mass content or past interactions with dark matter clumps.

The Ly$\alpha$ forest consists of absorption features in the spectra of distant quasars caused by Lyman-$\alpha$ transitions of neutral hydrogen in the foreground intergalactic medium (IGM). Thus, the Ly$\alpha$ forest is a biased tracer of the neutral hydrogen distribution, and hence, it probes the underlying dark matter density field. As a dark matter probe covering higher redshifts than traditional galaxy surveys, the forest has been used to measure the BAO feature; nonetheless, its relatively high redshift $2 \lesssim z \lesssim 5$ and sensitivity to small-scale clustering ($\gtrsim$ Mpc) make it an ideal probe of dark matter models\footnote{However, the Ly$\alpha$ forest is unlikely to detect subtle differences between dark matter models \cite{Valluri2022}, such as the oscillations predicted by FDM at very small scales \cite{Lague2021}.}. For instance, constraints from the 1D Ly$\alpha$ flux power spectrum, obtained from medium and high-resolution spectra, set a firm lower limit on the WDM particle mass of $m_{\rm X} \geq 5.3$ -- $5.7$ keV \cite{Palanque2020,2024PhRvD.109d3511I}.     

Although MW astrophysics and the Ly$\alpha$ forest can provide robust dark matter constraints, the accuracy of these constraints depends on careful modeling of baryonic physics and nonlinearities (e.g., \cite{Valluri2022}). In particular, Ly$\alpha$ forest constraints require precise modeling of IGM astrophysics \cite{Garzilli2021, Puchwein2023, Montero-Camacho2019b, Montero-Camacho2023} due to the degeneracies between the response of IGM to cosmic reionization and the impact of non-cold dark matter on the thermal state of the IGM. With a conservative flux limit of $r\leq 23.5$ (larger than that of DESI, which is $r<23$) and an estimated quasar number density surpassing 80 deg$^{-2}$ at $z > 2.1$, MUST is expected to refine our ability to constrain dark matter models using the Ly$\alpha$ forest (see Section~\ref{ssec:forecastwdm} and Figure~\ref{fig:wdm}).

\subsection{Synergy with Other Probes}
    \label{ssec:synergy}

\subsubsection{Imaging surveys}

Next-generation space-based wide-field imaging surveys, such as the China Space Station Telescope (CSST; \cite{Zhan2021}), Euclid \cite{Scaramella2022}, and Roman \cite{Akeson2019}, will measure galaxy shears with unprecedented precision, providing tighter constraints on dark energy through weak lensing. The ground-based Legacy Survey of Space and Time (LSST; \cite{LSST}) will observe the whole southern sky every few nights, building the most extensive catalog of transient phenomena of cosmological interest, such as type Ia supernovae or variable AGN. Other imaging surveys worth noting are the Hyper Suprime-Cam Subaru Strategic Program (HSC-SSP; \cite{2018PASJ...70S...4A, Aihara2022}) and the Mozi Wide Field Survey Telescope (WFST; \cite{Wang2023}), which will provide deeper broadband sky coverage than currently available data, as well as the Javalambre Physics of the Accelerating Universe Survey (J-PAS; \cite{Benitez2014}), a narrow-band imaging survey with higher-precision photometric redshifts.

In addition to using imaging data from these surveys for MUST target selection, we aim to maximize scientific outcomes through joint analyses of the data for cosmological measurements or through dedicated, coordinated observations.
For instance, the spectroscopic and photometric samples can be cross-correlated to break degeneracies associated with intrinsic alignments and galaxy bias, thus improving the precision of cosmological measurements (e.g., \cite{Joachimi2011, Tutusaus2020}).
Additionally, spare fibers from MUST can be used to complete the spectroscopic data of galaxies in color space, thereby expanding the training set for calibrating photometric or slitless spectroscopic redshift measurements (e.g., \cite{Myles2021, VanDenBusch2022}).
The high efficiency and large survey volume of MUST offer opportunities to identify or confirm strong lensing systems, e.g., low-$z$ bright galaxies with anomalously high-$z$ emission lines \cite{Treu2010}.
Moreover, the cross-correlation between imaging data and Ly$\alpha$ forests from MUST may enable tracing LSS using Ly$\alpha$ intensity mapping \cite{Renard2024}.

\subsubsection{CMB Experiments} 

The next generation of CMB experiments, such as AliCPT \cite{2017arXiv171003047L}, Simons Observatory \cite{2019JCAP...02..056A}, CMB-S4 \cite{2019arXiv190704473A}, LiteBIRD \cite{2024arXiv240602724G}, and PICO \cite{2018SPIE10698E..4FS}, hold significant promise for detecting primordial gravitational waves from the early-universe inflationary epoch. Precisely measuring primordial gravitational waves faces two significant obstacles: foreground contamination and gravitational lensing distortions. Consequently, accurate reconstruction of the CMB lensing potential is critical for isolating the primordial signal. Multi-tracer delensing methods, which leverage galaxy survey data, are expected to significantly enhance the capability of CMB experiments to detect the elusive primordial gravitational waves \cite{Namikawa2024}.

In addition to enhancing our understanding of inflationary physics, MUST will facilitate synergistic investigations of large-scale structures. This includes measuring the kinetic Sunyaev-Zel'dovich (kSZ) effect, which provides insights into the dynamics of baryonic matter (e.g., \cite{2018PhRvD..97b3514L}). Furthermore, cross-correlations with CMB lensing will improve our understanding of structure formation (e.g., \cite{2021MNRAS.501.6181K}). With these complementary approaches, MUST, in combination with next-generation CMB experiments, promises to enrich our understanding of cosmic evolution.

\subsubsection{Radio surveys}

The Square Kilometre Array (SKA; \cite{SKA2020}) will conduct an IM program using the 21\,cm hyperfine transition of neutral Hydrogen. Due to the significant disparity in strength between foreground emissions and the cosmological 21\,cm signal — differing by several orders of magnitude — confirming the cosmic origin of the signal, particularly at high redshifts ($z \gtrsim 1$), will likely require cross-correlations with other cosmological tracers. Current studies typically focus on cross-correlations with high-redshift galaxies \cite{Furlanetto2007, LaPlante2023}, LBG \cite{Villaescusa-Navarro2015}, and the Ly$\alpha$ forest \cite{Carucci2017, Montero-Camacho2025}. These complementary approaches, coupling 21\,cm measurements with spectroscopic surveys, enhance our ability to discern the cosmological signal from foreground contamination, thus improving the robustness of IM measurements. 

MUST will significantly enhance cross-correlation efforts involving 21\,cm observations of high-redshift galaxies and the Lyman-$\alpha$ forest, enabling high-signal-to-noise measurements. Such cross-correlations will be instrumental in advancing our understanding of various cosmological phenomena, such as dark energy constraints \cite{2021JCAP...02..016D}, modified gravity \cite{2023JApA...44....5D}, dark matter \cite{2019JCAP...12..058S}, and hydrogen reionization \cite{Montero-Camacho2025}. 

\subsubsection{Gravitational waves and Fast Radio Bursts}

Very high-energy, transient phenomena such as gravitational waves and Fast Radio Bursts (FRBs) may also provide valuable cosmological constraints. Ever since the first detection of a gravitational wave event by LIGO \cite{Aasi2015}, several large observatories are projected, e.g., LISA \cite{Amaro-Seoane2017}, Taiji \cite{Luo2021}, TianQin \cite{Luo2021}, and the Einstein Telescope \cite{Punturo2010}. In addition to the detection of singular events, the gravitational wave background has also been recently detected by Pulsar Timing Array collaborations (e.g., NANOGrav, \cite{Agazie2023}, CPTA \cite{Xu2023}, EPTA, \cite{Antoniadis2023}). Regarding FRBs, current radio observatories regularly detect these events, compiling ever-growing catalogs with hundreds of events (e.g., CHIME \cite{CHIMEFRB2021}, ASKAP \cite{Lu2019}, MeerKAT \cite{Rajwade2022}, and FAST \cite{Niu2021}).

The most direct synergy MUST will have is the measurement of redshifts and peculiar velocities for both the electromagnetic (EM) counterparts of gravitational wave sources and the FRB emitters. Gravitational waves from binary inspirals act as standard sirens since their luminosity distance can be inferred from their observation \cite{Holz2005}. Hence, a redshift measurement of its EM counterpart (either a transient event or a host galaxy) can provide an independent measurement of the Hubble constant \cite{Abbott2017}, and a large spectroscopic catalog of gravitational wave EM counterparts may yield independent $H_{\rm 0}$ measurements with a precision of a few percent \cite{Palmese2019}. In addition to the direct observation of gravitational wave EM counterparts, cross-correlation of gravitational wave events with spectroscopic galaxy catalogs may also yield improved joint constraints on $H_{\rm 0}$ \cite{Diaz2022, Borghi2024}.

Similarly, the distance to FRBs may also be determined by their dispersion measure \cite{You2007}. However, this method presents degeneracies with the baryon distribution in the FRB sightline, which in turn allows FRBs to constrain the baryon content of the Universe \cite{McQuinn2013, Huang2024}. Regardless of these degeneracies, measurements of $H_{\rm 0}$ using FRBs and spectroscopic redshifts of their host galaxies are possible \cite{James2022, Hagstotz2022}. Therefore, MUST will undoubtedly help with these measurements of $H_{\rm 0}$ by providing unprecedentedly large spectroscopic redshift catalogs of potential gravitational wave and FRB hosts.

\begin{table*}[!htb]
\centering
\resizebox{\textwidth}{!}{%
\begin{tabular}{|c|c|c|ccccccccc|c|}
\hline
\rowcolor[HTML]{EFEFEF} 
\cellcolor[HTML]{EFEFEF} &
  \cellcolor[HTML]{EFEFEF} &
  \cellcolor[HTML]{EFEFEF} &
  \multicolumn{9}{c|}{\cellcolor[HTML]{EFEFEF}\textbf{Imaging Depth (AB Mag)}} &
  \cellcolor[HTML]{EFEFEF} \\ \cline{4-12}
\rowcolor[HTML]{EFEFEF} 
\multirow{-2}{*}{\cellcolor[HTML]{EFEFEF}\textbf{Project}} &
  \multirow{-2}{*}{\cellcolor[HTML]{EFEFEF}\textbf{Survey Name}} &
  \multirow{-2}{*}{\cellcolor[HTML]{EFEFEF}\textbf{\begin{tabular}[c]{@{}c@{}}Survey Area \\ (deg$^2$)\end{tabular}}} &
  \multicolumn{1}{c|}{\cellcolor[HTML]{EFEFEF}{\color[HTML]{6434FC} \textbf{NUV}}} &
  \multicolumn{1}{c|}{\cellcolor[HTML]{EFEFEF}{\color[HTML]{3166FF} \textbf{$u$}}} &
  \multicolumn{1}{c|}{\cellcolor[HTML]{EFEFEF}{\color[HTML]{009901} \textbf{$g$}}} &
  \multicolumn{1}{c|}{\cellcolor[HTML]{EFEFEF}{\color[HTML]{F56B00} \textbf{$r$}}} &
  \multicolumn{1}{c|}{\cellcolor[HTML]{EFEFEF}{\color[HTML]{FE0000} \textbf{$i$}}} &
  \multicolumn{1}{c|}{\cellcolor[HTML]{EFEFEF}{\color[HTML]{963400} \textbf{$z$}}} &
  \multicolumn{1}{c|}{\cellcolor[HTML]{EFEFEF}{\color[HTML]{34696D} \textbf{Y}}} &
  \multicolumn{1}{c|}{\cellcolor[HTML]{EFEFEF}{\color[HTML]{303498} \textbf{J}}} &
  {\color[HTML]{646809} \textbf{H}} &
  \multirow{-2}{*}{\cellcolor[HTML]{EFEFEF}\textbf{Reference}} \\ \hline
\rowcolor[HTML]{DAE8FC} 
\cellcolor[HTML]{DAE8FC} &
  Optical Survey Wide &
  $\sim$ 17,500 &
  \multicolumn{1}{c|}{\cellcolor[HTML]{DAE8FC}{\color[HTML]{6434FC} 25.40}} &
  \multicolumn{1}{c|}{\cellcolor[HTML]{DAE8FC}{\color[HTML]{3166FF} 25.40}} &
  \multicolumn{1}{c|}{\cellcolor[HTML]{DAE8FC}{\color[HTML]{009901} 26.30}} &
  \multicolumn{1}{c|}{\cellcolor[HTML]{DAE8FC}{\color[HTML]{F56B00} 26.00}} &
  \multicolumn{1}{c|}{\cellcolor[HTML]{DAE8FC}{\color[HTML]{FE0000} 25.90}} &
  \multicolumn{1}{c|}{\cellcolor[HTML]{DAE8FC}{\color[HTML]{963400} 25.20}} &
  \multicolumn{1}{c|}{\cellcolor[HTML]{DAE8FC}{\color[HTML]{34696D} 24.40}} &
  \multicolumn{1}{c|}{\cellcolor[HTML]{DAE8FC}{\color[HTML]{303498} }} &
  {\color[HTML]{646809} } &
  \cellcolor[HTML]{DAE8FC} \\ \cline{2-12}
\rowcolor[HTML]{DAE8FC} 
\multirow{-2}{*}{\cellcolor[HTML]{DAE8FC}\textbf{CSST}} &
  Optical Survey Deep &
  $\sim$ 400 &
  \multicolumn{1}{c|}{\cellcolor[HTML]{DAE8FC}{\color[HTML]{6434FC} 26.70}} &
  \multicolumn{1}{c|}{\cellcolor[HTML]{DAE8FC}{\color[HTML]{3166FF} 26.70}} &
  \multicolumn{1}{c|}{\cellcolor[HTML]{DAE8FC}{\color[HTML]{009901} 27.50}} &
  \multicolumn{1}{c|}{\cellcolor[HTML]{DAE8FC}{\color[HTML]{F56B00} 27.20}} &
  \multicolumn{1}{c|}{\cellcolor[HTML]{DAE8FC}{\color[HTML]{FE0000} 27.00}} &
  \multicolumn{1}{c|}{\cellcolor[HTML]{DAE8FC}{\color[HTML]{963400} 26.50}} &
  \multicolumn{1}{c|}{\cellcolor[HTML]{DAE8FC}{\color[HTML]{34696D} 25.70}} &
  \multicolumn{1}{c|}{\cellcolor[HTML]{DAE8FC}{\color[HTML]{303498} }} &
  {\color[HTML]{646809} } &
  \multirow{-2}{*}{\cellcolor[HTML]{DAE8FC}\begin{tabular}[c]{@{}c@{}}Zhan et al. 2010; Cao et al. 2018; \\ Zhan et al. 2021\end{tabular}} \\ \hline
\rowcolor[HTML]{FFF3F3} 
\textbf{Euclid} &
  Wide Survey &
  $\sim$ 14,500 &
  \multicolumn{1}{c|}{\cellcolor[HTML]{FFF3F3}} &
  \multicolumn{1}{c|}{\cellcolor[HTML]{FFF3F3}{\color[HTML]{3166FF} }} &
  \multicolumn{1}{c|}{\cellcolor[HTML]{FFF3F3}{\color[HTML]{009901} }} &
  \multicolumn{3}{c|}{\cellcolor[HTML]{FFF3F3}{\color[HTML]{F56B00} $I_{\rm E}$ =26.20}} &
  \multicolumn{1}{c|}{\cellcolor[HTML]{FFF3F3}{\color[HTML]{34696D} 24.50}} &
  \multicolumn{1}{c|}{\cellcolor[HTML]{FFF3F3}{\color[HTML]{303498} 24.50}} &
  {\color[HTML]{646809} 24.50} &
  Scarmamella et al. 2022 \\ \hline
\rowcolor[HTML]{DAE8FC} 
\cellcolor[HTML]{DAE8FC} &
  Wide Fast Deep (WFD) &
  $\sim$ 14,500 &
  \multicolumn{1}{c|}{\cellcolor[HTML]{DAE8FC}} &
  \multicolumn{1}{c|}{\cellcolor[HTML]{DAE8FC}{\color[HTML]{3166FF} 25.30}} &
  \multicolumn{1}{c|}{\cellcolor[HTML]{DAE8FC}{\color[HTML]{009901} 26.84}} &
  \multicolumn{1}{c|}{\cellcolor[HTML]{DAE8FC}{\color[HTML]{F56B00} 27.04}} &
  \multicolumn{1}{c|}{\cellcolor[HTML]{DAE8FC}{\color[HTML]{FE0000} 26.35}} &
  \multicolumn{1}{c|}{\cellcolor[HTML]{DAE8FC}{\color[HTML]{963400} 25.22}} &
  \multicolumn{1}{c|}{\cellcolor[HTML]{DAE8FC}{\color[HTML]{34696D} 24.47}} &
  \multicolumn{1}{c|}{\cellcolor[HTML]{DAE8FC}{\color[HTML]{303498} }} &
  {\color[HTML]{646809} } &
  DESC et al. 2018 \\ \cline{2-13} 
\rowcolor[HTML]{DAE8FC} 
\multirow{-2}{*}{\cellcolor[HTML]{DAE8FC}\textbf{LSST}} &
  North Ecliptic Spur (NES) &
  $\sim$4,160 &
  \multicolumn{1}{c|}{\cellcolor[HTML]{DAE8FC}} &
  \multicolumn{1}{c|}{\cellcolor[HTML]{DAE8FC}} &
  \multicolumn{1}{c|}{\cellcolor[HTML]{DAE8FC}{\color[HTML]{009901} 25.64}} &
  \multicolumn{1}{c|}{\cellcolor[HTML]{DAE8FC}{\color[HTML]{F56B00} 25.84}} &
  \multicolumn{1}{c|}{\cellcolor[HTML]{DAE8FC}{\color[HTML]{FE0000} 25.15}} &
  \multicolumn{1}{c|}{\cellcolor[HTML]{DAE8FC}{\color[HTML]{643403} 24.02}} &
  \multicolumn{1}{c|}{\cellcolor[HTML]{DAE8FC}} &
  \multicolumn{1}{c|}{\cellcolor[HTML]{DAE8FC}} &
   &
  Assuming $1/3$ visits of WFD \\ \hline
\rowcolor[HTML]{FFF3F3} 
\cellcolor[HTML]{FFF3F3} &
  UNIONS/CFIS &
  $\sim$ 4,861 &
  \multicolumn{1}{c|}{\cellcolor[HTML]{FFF3F3}} &
  \multicolumn{1}{c|}{\cellcolor[HTML]{FFF3F3}{\color[HTML]{3166FF} 24.30}} &
  \multicolumn{1}{c|}{\cellcolor[HTML]{FFF3F3}{\color[HTML]{009901} 25.20}} &
  \multicolumn{1}{c|}{\cellcolor[HTML]{FFF3F3}{\color[HTML]{F56B00} 24.90}} &
  \multicolumn{1}{c|}{\cellcolor[HTML]{FFF3F3}{\color[HTML]{FE0000} 24.30}} &
  \multicolumn{1}{c|}{\cellcolor[HTML]{FFF3F3}{\color[HTML]{963400} 24.10}} &
  \multicolumn{1}{c|}{\cellcolor[HTML]{FFF3F3}{\color[HTML]{34696D} }} &
  \multicolumn{1}{c|}{\cellcolor[HTML]{FFF3F3}{\color[HTML]{303498} }} &
  {\color[HTML]{646809} } &
  \cellcolor[HTML]{FFF3F3} \\ \cline{2-12}
\rowcolor[HTML]{FFF3F3} 
\multirow{-2}{*}{\cellcolor[HTML]{FFF3F3}\textbf{UNIONS}} &
  Extended UNIONS $u$-band &
  $\sim$ 8,988 &
  \multicolumn{1}{c|}{\cellcolor[HTML]{FFF3F3}} &
  \multicolumn{1}{c|}{\cellcolor[HTML]{FFF3F3}{\color[HTML]{3166FF} 24.30}} &
  \multicolumn{1}{c|}{\cellcolor[HTML]{FFF3F3}{\color[HTML]{009901} }} &
  \multicolumn{1}{c|}{\cellcolor[HTML]{FFF3F3}{\color[HTML]{F56B00} }} &
  \multicolumn{1}{c|}{\cellcolor[HTML]{FFF3F3}{\color[HTML]{FE0000} }} &
  \multicolumn{1}{c|}{\cellcolor[HTML]{FFF3F3}{\color[HTML]{963400} }} &
  \multicolumn{1}{c|}{\cellcolor[HTML]{FFF3F3}{\color[HTML]{34696D} }} &
  \multicolumn{1}{c|}{\cellcolor[HTML]{FFF3F3}{\color[HTML]{303498} }} &
  {\color[HTML]{646809} } &
  \multirow{-2}{*}{\cellcolor[HTML]{FFF3F3}Ibata et al. 2017} \\ \hline
\rowcolor[HTML]{DAE8FC} 
\cellcolor[HTML]{DAE8FC} &
  DECaLS &
  $\sim$ 9,000 &
  \multicolumn{1}{c|}{\cellcolor[HTML]{DAE8FC}} &
  \multicolumn{1}{c|}{\cellcolor[HTML]{DAE8FC}{\color[HTML]{3166FF} }} &
  \multicolumn{1}{c|}{\cellcolor[HTML]{DAE8FC}{\color[HTML]{009901} 24.65}} &
  \multicolumn{1}{c|}{\cellcolor[HTML]{DAE8FC}{\color[HTML]{F56B00} 23.61}} &
  \multicolumn{1}{c|}{\cellcolor[HTML]{DAE8FC}{\color[HTML]{FE0000} 22.84}} &
  \multicolumn{1}{c|}{\cellcolor[HTML]{DAE8FC}{\color[HTML]{963400} }} &
  \multicolumn{1}{c|}{\cellcolor[HTML]{DAE8FC}{\color[HTML]{34696D} }} &
  \multicolumn{1}{c|}{\cellcolor[HTML]{DAE8FC}{\color[HTML]{303498} }} &
  {\color[HTML]{646809} } &
  \cellcolor[HTML]{DAE8FC} \\ \cline{2-12}
\rowcolor[HTML]{DAE8FC} 
\multirow{-2}{*}{\cellcolor[HTML]{DAE8FC}\textbf{Legacy Survey}} &
  BASS $+$ MzLS &
  $\sim$ 5,000 &
  \multicolumn{1}{c|}{\cellcolor[HTML]{DAE8FC}} &
  \multicolumn{1}{c|}{\cellcolor[HTML]{DAE8FC}{\color[HTML]{3166FF} }} &
  \multicolumn{1}{c|}{\cellcolor[HTML]{DAE8FC}{\color[HTML]{009901} 24.30}} &
  \multicolumn{1}{c|}{\cellcolor[HTML]{DAE8FC}{\color[HTML]{F56B00} 23.70}} &
  \multicolumn{1}{c|}{\cellcolor[HTML]{DAE8FC}{\color[HTML]{FE0000} 23.04}} &
  \multicolumn{1}{c|}{\cellcolor[HTML]{DAE8FC}{\color[HTML]{963400} }} &
  \multicolumn{1}{c|}{\cellcolor[HTML]{DAE8FC}{\color[HTML]{34696D} }} &
  \multicolumn{1}{c|}{\cellcolor[HTML]{DAE8FC}{\color[HTML]{303498} }} &
  {\color[HTML]{646809} } &
  \multirow{-2}{*}{\cellcolor[HTML]{DAE8FC}Dey et al. 2019} \\ \hline
\rowcolor[HTML]{FEE8E7} 
\textbf{Mozi} &
  Wide Field Survey &
  $\sim$8,000 &
  \multicolumn{1}{c|}{\cellcolor[HTML]{FEE8E7}} &
  \multicolumn{1}{c|}{\cellcolor[HTML]{FEE8E7}{\color[HTML]{3166FF} 24.82}} &
  \multicolumn{1}{c|}{\cellcolor[HTML]{FEE8E7}{\color[HTML]{009901} 25.85}} &
  \multicolumn{1}{c|}{\cellcolor[HTML]{FEE8E7}{\color[HTML]{F56B00} 25.36}} &
  \multicolumn{1}{c|}{\cellcolor[HTML]{FEE8E7}{\color[HTML]{FE0000} 24.83}} &
  \multicolumn{1}{c|}{\cellcolor[HTML]{FEE8E7}{\color[HTML]{963400} 23.90}} &
  \multicolumn{1}{c|}{\cellcolor[HTML]{FEE8E7}{\color[HTML]{34696D} }} &
  \multicolumn{1}{c|}{\cellcolor[HTML]{FEE8E7}{\color[HTML]{303498} }} &
  {\color[HTML]{646809} } &
  Wang et al. 2023 \\ \hline
\rowcolor[HTML]{DAE8FC} 
\cellcolor[HTML]{DAE8FC} &
  SSP Wide &
  $\sim$ 1,400 &
  \multicolumn{1}{c|}{\cellcolor[HTML]{DAE8FC}} &
  \multicolumn{1}{c|}{\cellcolor[HTML]{DAE8FC}{\color[HTML]{3166FF} }} &
  \multicolumn{1}{c|}{\cellcolor[HTML]{DAE8FC}{\color[HTML]{009901} 26.50}} &
  \multicolumn{1}{c|}{\cellcolor[HTML]{DAE8FC}{\color[HTML]{F56B00} 26.50}} &
  \multicolumn{1}{c|}{\cellcolor[HTML]{DAE8FC}{\color[HTML]{FE0000} 26.20}} &
  \multicolumn{1}{c|}{\cellcolor[HTML]{DAE8FC}{\color[HTML]{963400} 25.20}} &
  \multicolumn{1}{c|}{\cellcolor[HTML]{DAE8FC}{\color[HTML]{34696D} 24.40}} &
  \multicolumn{1}{c|}{\cellcolor[HTML]{DAE8FC}{\color[HTML]{303498} }} &
  {\color[HTML]{646809} } &
  \cellcolor[HTML]{DAE8FC} \\ \cline{2-12}
\rowcolor[HTML]{DAE8FC} 
\multirow{-2}{*}{\cellcolor[HTML]{DAE8FC}\textbf{HSC SSP}} &
  SSP Deep$+$UltraDeep &
  $\sim$ 37 &
  \multicolumn{1}{c|}{\cellcolor[HTML]{DAE8FC}} &
  \multicolumn{1}{c|}{\cellcolor[HTML]{DAE8FC}{\color[HTML]{3166FF} }} &
  \multicolumn{1}{c|}{\cellcolor[HTML]{DAE8FC}{\color[HTML]{009901} 27.40}} &
  \multicolumn{1}{c|}{\cellcolor[HTML]{DAE8FC}{\color[HTML]{F56B00} 27.10}} &
  \multicolumn{1}{c|}{\cellcolor[HTML]{DAE8FC}{\color[HTML]{FE0000} 26.90}} &
  \multicolumn{1}{c|}{\cellcolor[HTML]{DAE8FC}{\color[HTML]{963400} 26.30}} &
  \multicolumn{1}{c|}{\cellcolor[HTML]{DAE8FC}{\color[HTML]{34696D} 25.30}} &
  \multicolumn{1}{c|}{\cellcolor[HTML]{DAE8FC}{\color[HTML]{303498} }} &
  {\color[HTML]{646809} } &
  \multirow{-2}{*}{\cellcolor[HTML]{DAE8FC}Aihara et al. 2022} \\ \hline
\end{tabular}%
}
\caption{
    Summary of available, ongoing, and planned imaging surveys that overlap with the potential footprint of MUST and can contribute to the target selection for LSS tracers of MUST. Please see the reference for details. The imaging depths correspond to the 5\,$\sigma$ point-source detection limits. However, we ignore the minor differences in aperture choices and filter differences between surveys. Please see Figure~\ref{fig:surveys} to visualize their footprints.
}
\label{table2}
\end{table*}

\section{Target Selection}
    \label{sec:target}

Target selection is the bridge between imaging data and spectroscopic measurements, and is critical for achieving the science goals outlined in Section~\ref{sec:motivation}. With finite observational resources, there are two main requirements for cosmological target selection:
1) There should be clear spectral features, such as strong emission lines, absorption features, or continuum breaks, to secure redshift determination in short exposures; and
2) It should reach a target density that balances cosmic variance and shot noise to maximize the precision of cosmological measurements.
Different science goals emphasize different scales, redshift ranges, and number densities, so trade-offs are inevitable.
Ideally, one would explore the full target-selection space for all science cases, including topics beyond large-scale structures, and then jointly optimize the survey strategy.

Given the current state of available imaging samples, a full exploration is impractical at the moment. In this section, we therefore adopt simple extensions of Stage-IV (e.g., DESI-like) selections and test how far MUST can go on the major cosmological goals, such as dark energy and modified gravity, within a realistic survey duration (see Section~\ref{sec:forecast}).
As deeper and wider imaging data with more bands become available, we will broaden the selection methods and perform a more complete optimization across all science cases in future work.


\subsection{Challenges of Target Selection for Stage-V Spectroscopic Surveys}
    \label{ssec:targetchallenge}

Unlike spectroscopic surveys using a prism (e.g., PRIMUS, \cite{PRIMUSI, PRIMUSII}), grism (e.g., 3D-HST, \cite{3DHST}; Euclid \cite{Euclidcoll2024}), or IFS (e.g., HETDEX \cite{HETDEX}), modern surveys using multi-slit instruments or robotic fiber positioners require careful and sophisticated target selection based on multi-band imaging data.  Starting with the Main Galaxy Sample (MGS) in SDSS, a primarily flux-limited sample, subsequent surveys have gradually expanded the number of samples. As a Stage-IV survey, the main survey of DESI has already included low-redshift bright galaxies (BGS; \cite{DESI_BGS}), luminous red galaxies (LRG; \cite{DESI_LRG}), emission-line galaxies (ELG; \cite{DESI_ELG}), and quasi-stellar objects (QSO; \cite{DESI_QSO}). Selecting these samples now involves more complex criteria, including multiple color cuts, cross-matching with multi-wavelength datasets, and the application of machine-learning methods (e.g., \cite{Hoyle16, DarraghFord23}). 

These changes reflect the increasingly demanding scientific and operational requirements of modern surveys. A survey must design flexible programs for bright \& dark nights and different observing conditions while maximizing the fiber efficiency and the scientific output of the project. More importantly, for LSS redshift surveys, target selection design needs to ensure the volume densities, redshift distributions, and average halo biases of different LSS tracers meet the requirements for constraining the cosmological model. At the same time, the target selection should intentionally minimize the systematics inherited from imaging surveys, such as target density fluctuations induced by variations in imaging depth, observing conditions, Galactic extinction, and data reduction. As the number of redshifts rapidly increases, the Stage-IV spectroscopic surveys have officially entered the low-shot-noise regime, where systematic issues arising from target selection are becoming critical to the survey's cosmological potential. Looking forward to the Stage-V era for MUST, we anticipate not only the continued development of these trends but also new challenges.

In particular, the LSS survey of MUST faces two outstanding challenges. First, the high-redshift ($2<z<5$) surveys of Stage V require new LSS tracers that differ from those adopted by Stage III and Stage IV projects, such as ELGs and LRGs. The best candidates are Lyman Break Galaxies (LBG) selected using the Lyman-break ``dropout'' technique and the Lyman-$\alpha$ Emitters (LAEs) traditionally identified in narrow-band imaging surveys. These types of high-$z$ populations, representing galaxies with a wide range of stellar mass, star-formation rate, and halo properties (e.g., halo bias), should be able to continuously cover the redshift range observable by the spectrograph of MUST ($2<z<5$) with sufficient tracer densities (e.g., \cite{Im24}). However, while the high-$z$ galaxy community has been studying these populations extensively for two decades now (e.g., Hu \& McMahon 1996 \cite{Hu96}; Cowie \& Hu 1998 \cite{Cowie98}), we still have not acquired the demanding broad- and narrow-band imaging capabilities to select them for a Stage-V survey and also have not fully understood their cosmological potential as existing deep field studies are still limited by cosmic variance among other systematics. As the LSS survey community has begun to pay attention and organize pilot programs (e.g., \cite{Ruhlmann-Kleider2024, White2024}), we expect to gain a better understanding of these new LSS tracers before MUST first lights. The proposed DESI-II project is exciting, as it could launch the first LSS survey using LBG and LAEs in $\sim$2030, providing valuable insights before MUST's operation.

Secondly, given the site selection of MUST in the northern hemisphere, we do not enjoy the deep \& uniform multi-band imaging coverage of the Legacy Survey of Space and Time (LSST) of the Vera C. Rubin Observatory in the south. Based on the current landscape of imaging surveys in the north, no similar survey is expected to be available before the 2030s. The {\it left} panel of Figure~\ref{fig:surveys} visualizes the footprints of major imaging surveys that overlap with the potential footprint of MUST and can contribute to the target selection of MUST. We also summarize their survey areas and imaging depths in Table~\ref{table2}. Assuming MUST will observe targets with airmass better than 1.5, at the Lenghu site, MUST can cover the $\delta \geq 10^{\circ}$ footprint. Within this area, no imaging survey can cover the high Galactic latitude region suitable for an LSS survey. In the ideal scenario, the Chinese Space Station Telescope's optical survey (CSST-OS) will provide the best multi-band support from the NUV to the $Y$-band. The NUV and $u$-band observations are particularly important for selecting $z>2$ LBG. Meanwhile, the {\it Euclid} mission will provide valuable near-infrared (NIR) coverage and deep observations in the broad $I_{\rm E}$ optical band. While the exact strategy has yet to be investigated, the synergy between optical and NIR images from space is expected to provide unique advantages and new perspectives for target selection. However, based on the current survey plan, both CSST and {\it Euclid} will avoid the region near the Ecliptic plane. For this region, the LSST survey should provide partial coverage, but most of it was covered by the Northern Ecliptic Spur (NES) mini-survey\footnote{\url{https://survey-strategy.lsst.io/baseline/minis.html}} of (\cite{Schwamb18}), which will only observe in $g$, $r$, $i$, \& $z$-filters with $\sim 1/3$ of the visits of the main Wide-Fast-Deep (WFD) survey. While the coadding depth is still competitive, the lack of the $u$-band is unfortunate for the LBG selection. The ongoing Ultraviolet Near Infrared Optical Northern Survey (UNIONS) campaign using the 3.6 m CFHT, 2$\times$ 1.8 m Pan-STARRS, and 8.2 m Subaru telescopes is another major candidate for multi-band target selection. In an ideal situation, the southern extension of UNIONS in the $u$-band to the northern limit ($\delta = +12^{\circ}$) of the main LSST survey will be especially valuable. 

In addition to these ongoing or planned imaging surveys, we expect the available multi-band data from the Legacy Survey (in $g$, $r$, and $z$-bands; \cite{Dey2019}) and the Hyper Suprime-Cam Subaru Strategic Program (HSC-SSP; \cite{2018PASJ...70S...4A}) will be essential for the development of target selection strategy for MUST: Legacy Survey satisfies the survey footprint requirements for a northern Stage-V project, while HSC-SSP, along with the UNIONS $u$-band data should meet or surpass the imaging depth requirements for MUST. Additionally, the 2.5 m Mozi wide-field survey telescope at Peak C of the Lenghu site has commenced operation. As a dedicated time-domain survey telescope that will continue its operation into the 2030s, its accumulated imaging depths will also be helpful to MUST. Moreover, the potential to image the northern sky with customized narrow- or medium-band filters to facilitate LAE selection is an intriguing opportunity to be explored. We should also mention that CFHT, Subaru, and the 4.0 m Blanco telescope still have wide-field imaging capability that could support imaging campaigns to provide targeting selection support for any Stage-V spectroscopic surveys. In particular, the narrow- and medium-band surveys using the Dark Energy Camera (DECam; \cite{DECam}), such as the One-hundred-deg$^2$ DECam Imaging in Narrowbands (ODIN; \cite{Lee2023}), the Merian survey (e.g., \cite{Merian}), and J-PAS (e.g., \cite{Benitez2014}) could demonstrate an interesting new approach for target selection for the Stage-V survey.

Put all together, we can see that MUST is facing a highly challenging task: developing a consistent target-selection strategy that meets the strict requirements for a Stage-V survey. As this work focuses on the theoretical prediction of LSS cosmology, we will not delve into the details of target selection, as many of the imaging datasets mentioned above are not yet available. Instead, in the following subsections, we will briefly introduce the concepts and assumptions of target selection for the LSS survey of MUST. 

For the cosmological prediction, we will assume two survey footprint scenarios: an ideal gray (dark) time survey with a 13,000\,deg$^2$ (11,000\,deg$^2$) footprint, and a more conservative 11,000\,deg$^2$ (8,000\,deg$^2$) footprint. 

\begin{figure*}[!htb]
    \centering
    \includegraphics[width=1.0\linewidth]{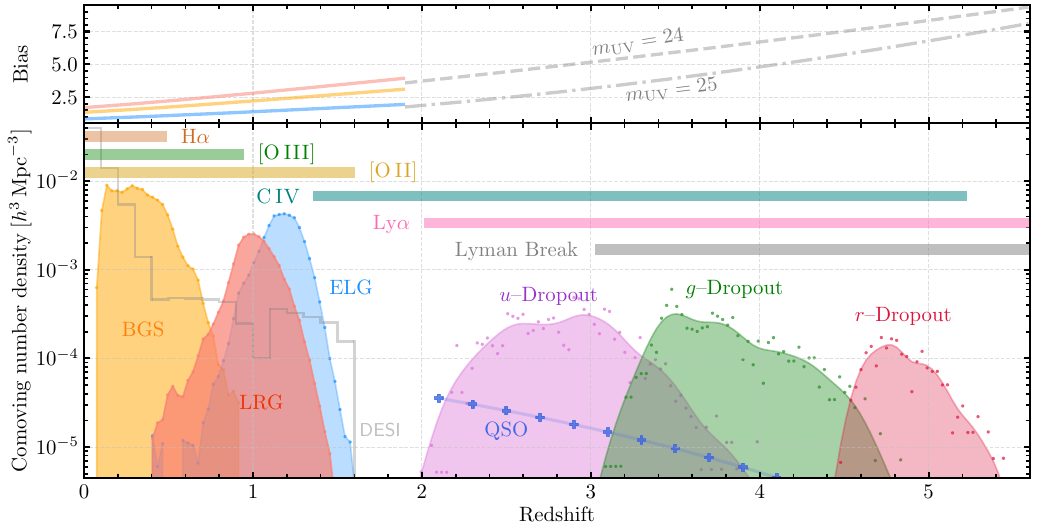}
    \caption{
        Preliminary target selection strategy for the LSS spectroscopic surveys of MUST.
        {\it Bottom} panel: redshift distributions and comoving number densities of the primary LSS tracers targeted by MUST at low--intermediate redshift ($z \lesssim 1.6$) and in the high-redshift Universe. The expected DESI number densities used for the forecasts in this work are also shown. At $z \lesssim 1.5$, MUST will densely sample the LSS using the low-redshift bright galaxy sample (BGS), luminous red galaxies (LRGs), and emission line galaxies (ELGs), with their [\ion{O}{II}] doublets, other prominent optical emission lines ([\ion{O}{III}], H$_{\beta}$, H$_{\alpha}$, etc.), and significant absorption features. Horizontal colored bars indicate the wavelength coverage of the MUST spectrograph for different emission lines. At $z \gtrsim 2$, MUST will use the dropout technique to select Lyman-break galaxies (LBGs) over a large comoving volume to map the high-$z$ LSS with unprecedented detail. Based on expected imaging data, we show initial target selection results for $u$-, $g$-, and $r$-dropout galaxies. For these LBGs, redshifts will be measured using the Lyman break feature, the Lyman-$\alpha$ emission line, and strong interstellar absorption lines such as \ion{C}{IV}. MUST will also observe quasi-stellar objects (QSOs) at $z \gtrsim 2$ as complementary high-redshift LSS tracers. 
        {\it Top} panel: redshift evolution models for the biases of BGS, LRGs, and ELGs at $z \gtrsim 2$, along with a bias model for LBGs with different UV magnitude limits. These models serve as the basis for the cosmological forecast presented in this work. 
    }
    \label{fig:nz_bias}
\end{figure*}

\subsection{Low-Redshift Tracers}
    \label{ssec:lzg}

Similar to the ongoing DESI survey, MUST can measure the redshift of $z<1.57$ galaxies using the [\ion{O}{II}] doublet emission lines at 3727 \& 3729\,\AA{}, along with the other significant emission and absorption features in galaxies' spectra. While this ``low-redshift'' component is not the most critical one for the cosmological goals of MUST, it will be a valuable dataset that enables many other complementary cosmological probes, such as the multi-tracer clustering to explore the non-linear regime (e.g., \cite{Bull2017, Favole2019}), the cross-correlations with weak lensing surveys like CSST \& {\it Euclid} and other multi-wavelength datasets (e.g., \cite{Joachimi2021, Kids1000_2021, Lange2023}), and the spectroscopic survey of galaxy clusters (e.g., \cite{Sohn2021, Sohn2023}). 

For now, we assume that MUST will follow the recipes of DESI to select BGS \cite{DESI_BGS}, LRG \cite{DESI_LRG}, and ELG \cite{DESI_ELG} samples as $z<1.6$ LSS tracers. Conceptually, we plan to observe BGS and LRG samples as bright-night targets, while observing ELG during gray and dark nights. 

By 2030, DESI and its extension should have finished collecting more than 40 million redshifts in this range. Therefore, we will focus only on targets fainter than those in DESI. Although the methods are similar, the BGS, LRG, and ELG samples for MUST were selected using photometric data from Legacy Survey DR10.1, rather than DR9. As Legacy Survey DR11 is scheduled for release in 2025, it would be interesting to check if the samples can be improved with this new release.

\subsubsection{Bright Galaxy Sample (BGS)}
    \label{ssec:bgs}

MUST aims to extend the BGS sample to the fainter end, from $r < 20.175$\,mag (including the {\tt FAINT} sub-sample) to $r < 21$\,mag, and to secure a 2,000 deg$^{-2}$ target density. Using the recipe established by DESI \cite{DESI_BGS}, we apply the same quality cleaning cuts and spatial mask bits (\texttt{BITMASK} 1, 12, and 13) to the LegacySurvey objects that have been observed at least once in the $g, r $ and $z$ bands, exclude stellar objects using the {\it Gaia} catalog, and also remove spurious objects with extreme colors. As we only require rough target density and redshift distribution estimates for cosmological forecasts, we have not accounted for the different fiber sizes between MUST and DESI.

In Figure~\ref{fig:nz_bias}, we show that the redshift distribution of the BGS sample ranges between $0.1 < z < 0.8$ with a broad peak between $z=0.2$ and $z=0.4$. For the halo bias assumption for the BGS, as this work intends to compare with the forecast of DESI, we adopt the same bias evolution model used in the scientific requirement document of DESI \cite{DESIcoll2016}:
\begin{equation}
    b_{\rm BGS} (z) = 1.34 / D(z) ,
\end{equation}
where $D(z)$ is the linear growth factor that depends on the adopted cosmology \cite{Heath1977}. 

In addition, by reaching $r <21.5$\,mag, MUST can increase the target density for BGS by $\sim$2,500\,deg$^{-2}$. Even considering the fiber assignment efficiency and the redshift success rate, which should be higher than 95\% given the fiducial exposure time, MUST can easily provide a dense sample of low-$z$ galaxies for various cosmological probes.   

\subsubsection{Luminous Red Galaxies (LRG)}
    \label{ssec:lrg}

As their name suggests, LRG represents the massive and (mostly) quiescent galaxies at low redshift ($z \lesssim 0.8$). LRG typically resides in relatively massive dark matter halos with high LSS bias, making them promising candidates for tracing the matter distribution and providing high-quality BAO measurements (e.g., \cite{Eisenstein2005, Alam2017}). 
Similar to the BGS sample, we explore a conceptual LRG selection using the LegacySurvey data, including the 3.4\,$\mu$m  $W1$ infrared band from the NEOWISE (e.g., \cite{NEOWISE}) data, and the DESI criteria \cite{DESI_LRG}. We exclude possible stars with the help of the {\it Gaia} EDR3 catalogs \cite{Gaia_EDR3} and apply the \texttt{BITMASK} 1, 12, and 13 bit masks. In the end, MUST can achieve a 2,000\,deg$^{-2}$ LRG density by extending the $z$-band fiber magnitude from $z_\mathrm{fiber} < 21.6$ mag to $z_\mathrm{fiber} < 22.8$ mag, reaching the detection limit of LegacySurvey. The current selection also ignores the fiber-size difference and requires that all objects have $g$, $r$, $z$, and $W1$ observations. 

Figure~\ref{fig:nz_bias} visualizes the target density and redshift distribution of our LRG selection, which ranges between $0.4 <z < 1.4$ and peaks at $z\sim 1.0$. For the bias evolution model of the LRG, we adopt the DESI one \cite{DESIcoll2016}:
\begin{equation}
    b_{\rm LRG} (z) = 1.7 / D(z).
\end{equation}

Note that the current LegacySurvey data can also provide additional LRGs with a few hundred per deg$^2$ density and even higher average redshift value down to $z_\mathrm{fiber} < 23.0$ if completeness is not a critical issue. Also, in the near future, we will explore the benefit of using deeper images from CSST, LSST, and Euclid to select LRGs for MUST's cosmology surveys.


\subsubsection{Emission Line Galaxies (ELG)}
    \label{ssec:elg}

While their galaxy-halo connection model (e.g., \cite{Yuan2023, Gao2024}) is still a significant systematic for today's LSS survey, the abundant population of ELGs at $z<1.6$ is the most appealing LSS tracer in the Stage-III and IV eras (e.g., \cite{Raichoor2017, DESI_ELG}). For a Stage-V survey such as MUST, high-density ELG samples with fainter magnitude limit enable a "high-fidelity'' map of LSS at $z\sim 1$ with interesting cosmological potential. Following DESI's strategy (color cuts, quality cuts, spatial mask, and clean photometry criteria) on LegacySurvey data again \cite{DESI_ELG}, we can select a conservative (optimistic) ELG sample with a 2,500 (3,600)\,deg$^{-2}$ target density by extending the $g$-band fiber magnitude limit from DESI's $24.1$\,mag to $24.6$\,mag, which will increase the DESI {\tt ELG\_LOP} sample's target density by 1.3 to 1.9 times.

We show the redshift distribution of ELG and volume density in Figure~\ref{fig:nz_bias}. Since we copied the ELG selection recipe from DESI, it is unsurprising that this sample spans the same $0.6 < z < 1.6$ redshift range and peaks at $z \sim 1.2$.  We adopt the same bias model as DESI \cite{DESIcoll2016}

\begin{equation}
    b_{\rm ELG} (z) = 0.84 / D(z) .
\end{equation}

It is worth noting that while the density of ELG is significantly higher than that of DESI, our current target selection is still based on the relatively shallow Legacy Survey data, which was designed for Stage-IV surveys. Such a selection is not optimized to isolate the targets at higher redshift, which should be the main focus of MUST. Deeper imaging data will help. Also, given the anticipated decrease of spectrograph throughput at $\lambda < 9000$\,\AA{}, the redshift success rate for faint ELG at the higher end of their redshift distribution strongly depends on the EW of the [\ion{O}{II}] emission lines. We will search for improved color selections to isolate the strong-[\ion{O}{II}] emitters at $z\gtrsim 1$. 

\subsection{High-Redshift Tracers}
    \label{ssec:hzg}

Starting from $z=2.1$, the Lyman-$\alpha$ emission line at rest-frame 1216\,\AA{} redshifts into the wavelength coverage at $\sim 370$\,nm for the spectrograph of MUST and enables our high-redshift target selection for the Stage-V cosmological survey.  Along with the Lyman break feature at rest-frame 912\,\AA{} and a series of interstellar absorption lines, such as \ion{C}{II} 1335\,\AA{}, \ion{C}{IV} 1548\,\AA{}, \ion{O}{I} 1302\,\AA{}, \ion{Si}{II} 1304\,\AA{}, and more (e.g., \cite{Shapley2003, Ruhlmann-Kleider2024}), support the redshift measurements at $2.1 < z < 5.0$ of MUST. High-$z$ LSS tracers are the bread and butter of a Stage-V cosmological survey, such as MUST. Until the Stage-IV surveys such as DESI, high-$z$ QSOs are the primary LSS tracers at $z>2$ (e.g., \cite{LeGoff2011, duMasdesBourboux2017, Ravoux2020, duMasdesBourboux2020}). While these accreting super-massive black holes (SMBHs) present us the unique opportunity to explore the intergalactic medium (IGM) to constrain the nature of dark matter (e.g., \cite{Garzilli2021, Villasenor2023, Fuss2023}), their volume density is typically too low to become the principal LSS tracer in the Stage-V era.

Currently, the most promising candidates for $z>2$ LSS tracers are the LBGs and LAEs (e.g., \cite{Wilson2019, White2024, Ruhlmann-Kleider2024, Im24}). Compared to LAEs, selecting LBGs based on broadband color criteria makes it easier to continuously populate the $2<z<5$ redshift space with sufficient density. Given the available data and previous work, we focus on the LBG populations as the primary LSS tracers for our cosmological forecast. While not included in the Fisher forecast, we briefly discuss the potential for LAEs and QSOs. 

We should note that LBGs and LAEs are defined photometrically, not physically, and that there is overlap between these two populations. In fact, the success rate of redshift measurement for LBGs strongly depends on the presence of a prominent Lyman-$\alpha$ emission line, which often makes them LAEs. 

\subsubsection{Lyman-Break Galaxies (LBG)}
    \label{ssec:lbg}

LBG are galaxies with significant flux decrement, or a ``break'', at the rest-frame wavelength shorter than the Lyman limit (911.3\,\AA{}) due to strong internal absorption by neutral Hydrogen. They represent the ``normal'', young, star-forming high-redshift galaxies (e.g., \cite{Steidel1996, Steidel1999, Reddy2009, Giavalisco2002}) with a large enough density to support LSS surveys. In observation, this break often translates into the non-detection in filters bluer than the observed frame of the Lyman limit (or a ``dropout'') or a high color value using filters that place the break between them. Starting from \cite{Steidel1996}, this method and its further development have helped select millions of LBG as candidates of high-$z$ galaxies in $2<z<7$ (e.g., $\sim 4.1$ million from the Great Optically Luminous Dropout Research Using Subaru HSC, or GOLDRUSH project \cite{Ono2018}).
This mature approach is the foundation for selecting high-$z$ LSS tracers using deep broadband imaging surveys for Stage-IV and Stage-V surveys. 

Motivated by the target selection for the proposed DESI-II, \cite{Ruhlmann-Kleider2024} validated the LBG selections based on a series of dropout criteria from the HSC SSP and the CLAUDS data using dedicated DESI campaigns and confirmed a $\sim 620$\,deg$^{-2}$ $r < 24.2$\,mag LBG density at $2.3 <z < 3.5$ or a $\sim 470$\,deg$^{-2}$ $r < 24.5$\,mag LBG density at $2.8 <z < 3.5$.  Also aimed at the fixed 1,100\,deg$^{-2}$ target density for DESI-II, \cite{Payerne2024} applies a Random Forest selection to the HSC$+$CLAUDS data and spectroscopically confirms a 493\,deg$^{-2}$ $z>2.5$ under the imaging depth afforded by a UNIONS-like survey.  

In \cite{Wilson2019}, the authors considered the BX color selection in \cite{Reddy2008} centered at $z\sim 2.20$, the $u$-dropout selection based on the CFHTLS-Archive-Research-Survey (CARS; \cite{Hildebrandt2009, Cooke2013}) centered at $z\sim 2.96$, and the $g$- \& $r$-dropout selections based on GOLDRUSH centered at $z\sim 3.8$ \& 4.9. Based on imaging depths of LSST-Y10, the authors concluded that it is practical to spectroscopically confirm 2,000 $R<24.0$\,mag BX galaxies, 500 $i<24.0$\,mag $u$-dropouts, 330 $i<25.5$\,mag $g$-dropouts, and 100 $z<25.5$\,mag $r$-dropouts per square degree. Altogether, this supports $\sim$ 3,000\,deg$^{-2}$ \emph{confirmed} high-$z$ tracers that match MUST's multiplex capability nicely.  Interestingly, the authors note that the $u$-band imaging depth limits the photometric selection of $ u$-dropouts. This assumes that one needs an actual detection in the $u$-band (or the band on the bluer side of the observed frame Lyman limit) to estimate the $u-g$ color to identify the break reliably, which is different from the dropout selections used to study the LBG populations (e.g., their luminosity function). It is worth investigating whether this is required to select LSS tracers with straightforward systematics. This is particularly important for MUST, as heterogeneous imaging datasets will contribute to our target selection. In this work, we still assume that detection is optional in the band $X$ to select the $X$-dropout populations. 

In the WST Science White Paper \cite{Mainieri2024}, the authors estimated the relations between the surface densities of $u$-, $g$-, and $r$-dropouts down to different magnitude limits (see Figure 65): at $r_{\rm lim} < 24.5$\,mag, the available \emph{photometric} density of $u$-dropout should be within 1,500 to 2,000\,deg$^{-2}$ for MUST. And, at $i_{\rm lim} < 24.6$, there are $\sim$1,000\,deg$^{-2}$ $g$-dropout for MUST to target. For $r$-dropout, the expected candidate density is $\sim$ 300 (600)\,deg$^{-2}$ for a $z_{\rm lim} < 24.5$ (25.0)\,mag sample. These values represent the most optimistic estimations for MUST. In reality, the photometric selection will include contamination from galaxies outside of the desired redshift range. MUST will not be able to measure the redshifts of all candidates as the success rate should strongly depend on the presence and the strength of the Ly$\alpha$ emission line. 

For the cosmological forecast in this work, we will not delve into the details of target selection, as we do not yet have the multi-band images and spectroscopic validation data required for a definitive Stage-V selection strategy. Instead, based on previous work, we provide optimistic and conservative estimates of the number density for each population. For $u$-dropout in $2.1 <z <3.5$, we assume that a $r<24.5$ selection could result in a 1,200\,deg$^{-2}$ sample in an optimistic and a 600\,deg$^{-2}$ sample in a conservative situation. For $g$-dropout in $3.3 <z < 4.5$, we expect a 800\,deg$^{-2}$ and 300\,deg$^{-2}$ sample for the optimistic and conservative scenarios, respectively. As for the $r$-dropout in $4.5 < z < 5.5$, we assume the optimistic sample has a density of $ 200\,\mathrm{deg}^{-2}$, while the conservative sample halves this value. We summarize these assumptions in Table~\ref{table2}. We do not include the BX selection sample in our forecast as it was defined using a different series of broadband filters than the Sloan $ugriz$ ones \cite{Gunn1998}. Hence, the BX criteria require filter conversions to work on the imaging data for MUST, or need to be updated. Moreover, the BX-selected sample often centers at $z\sim 2$, resulting in a significant fraction that falls outside the MUST-observed redshift range. Still, the current estimate suggests that the potential LSS tracer density at $z < 2.5$ could be much higher with better selection criteria. 

To provide the initial estimates of the redshift distributions for each population, we adopt a simple method based on the COSMOS2020\footnote{\url{https://astroweaver.github.io/project/cosmos2020-galaxy-catalog/}} photometric and photo-$z$ catalog \cite{Weaver2022}. Following the previous dropout-selection recipes, we designed the color cuts for the magnitude-limited MUST samples to roughly achieve the desired target density while ensuring a photometric redshift distribution consistent with the design. We apply the bright object masks from the COSMOS field of the HSC-SSP Ultra-Deep survey. It results in an effective footprint of 1.5\,deg$^2$ to estimate the target density. We confirm that the choice of photometric measurements and the photo-$z$ estimations do not affect the derived redshift distribution. 

For the $u$-dropouts, we recover a $\sim$1,500\,deg$^{-2}$ sample within $22.4 < r < 24.5$~mag using the following criteria: 
\begin{equation}
    \begin{aligned}
        & (u-g) > 0.95 ,\\
        & -0.5 < (g-r) < 1.1 ,\\
		& (u-g) > 1.17 \;\times\; (g-r) + 0.71.
    \end{aligned}
\end{equation}

About 88\% of the target has photo-$z$ within $2.2<z<3.5$ and shows a clear peak at $z=3$. We confirm that, while using different choices of $ugr$ or $ugi$ color cuts can change the target density from $\sim$800 to $>$2,000~deg$^{-2}$, the redshift distribution does not vary significantly. 

For the $g$-dropouts, we design the following $grz$ color cuts to select a $\sim$ 870~deg$^{-2}$ sample peaked at $z\sim 3.4$. $\sim 84$\% of the sample falls into $3.2 <z<4.5$ based on photo-$z$, resulting in a $\sim$ 760~deg$^{-2}$  sample within $23.0 < i < 24.6$~mag: 
\begin{equation}
    \begin{aligned}
        & (g-r) > 1.0 ,\\
		& (g-r) >  1.2 \times (r-z) + 0.65 .
    \end{aligned}
\end{equation}

As for the $r$-dropouts, we can define a $23.0 < z < 25.0$~mag sample within $4.0 < z < 5.5$, centered at $z\sim 4.5$ using these $riz$ color cuts: 
\begin{equation}
    \begin{aligned}
        & (r-i) > 0.65 ,\\
        & -0.5 < (i-z) < 0.9 ,\\
		& (r-i) >  1.5 \times (i-z) + 0.65 .
    \end{aligned}
\end{equation}
Using photo-$z$, this sample has a $\sim 390$~deg$^{-2}$ density within the desired redshift range. Like the $u$-dropouts, tweaking these preliminary color cuts can significantly change the target density but will not affect the redshift distributions.

We want to emphasize that these color cuts are \emph{not} designed for actual target selection, but only to support our \emph{assumptions} for a more optimistic high-$z$ LSS tracer selection in the 2030s.
For the clustering forecasts, we adopt the bias model of $ugr$-dropout LBGs in \cite{Wilson2019}:
\begin{equation}
b_{\rm LBG} (z, m) = A(m) (1 + z) + B(m) (1 + z)^2,
\end{equation}
where $A(m) = -0.98 (m - 25) + 0.11$ and $B(m) = 0.12 (m - 25) + 0.17$, with $m$ being the rest-frame UV magnitude.

LBG will be the highest-priority target for MUST's dark time program. However, based on the empirical estimation of the required exposure time (e.g., \cite{Wilson2019, Mainieri2024}) for LBG and the results from recent DESI campaigns (e.g., \cite{Ruhlmann-Kleider2024, Payerne2024}), successfully recovering redshifts down to $r\sim 24.5$\,mag or deeper is still a challenging task for a 6.5~m telescope. Not only does this challenge require a careful review of scientific requirements to guide the design of the scientific instruments, but it also motivates us to investigate an improved selection strategy that focuses on LBG with a clear Ly$\alpha$ emission line. As the throughput of the telescope, fiber, and detector all drop significantly toward the blue end at $\lambda < 4500$\,\AA{}, we expect the redshift success rate for $u$-dropout to also reflect this at $z < 2.5$. Given that we have not yet included the BX selection with a high target density, in principle, MUST should be able to guarantee the 3D tracer density for cosmology at $z < 2.5$. More importantly, despite uncertainties in the fiber assignment and redshift efficiency for LBG, the contrast between our optimistic and conservative target density estimates should make our cosmological forecast more realistic for MUST. 

\subsubsection{\texorpdfstring{Lyman-$\alpha$ Emitters (LAE)}{Lyman alpha Emitters}}
    \label{ssec:lae}

LAEs are high-$z$ galaxies with strong (${\rm EW} \geq 20$\,\AA{}) Lyman-$\alpha$ emission lines (e.g., \cite{Ouchi2020}), which are typically young, star-forming galaxies with low stellar and dust mass. As LAEs live in dark matter halos with a lower average halo mass than LBG, their potential density should be significant enough to make them competitive high-redshift LSS tracers. In \cite{White2024}, the authors spectroscopically confirm 822 (1,099) $z=2.40\pm0.03$ ($3.10\pm0.03$) LAEs in the 8.90 (9.34)\,deg$^2$ ODIN fields using the N419 (N501) narrow-band filters down to 25.5 (25.7) 5\,$\sigma$ detection limit. These samples correspond to a $\sim 10^{-3}\,h^{-1}\,{\rm Mpc}^{-3}$ 3D density in two narrow redshift windows, higher than the optimistic prediction of LBG 3D density across the whole redshift range. If such performance can be extrapolated to a broader redshift range, LAEs can become an extremely interesting LSS tracer: not only do they dramatically improve the density of high-$z$ LSS tracers, but their halo bias should also be significantly lower than the dropout-selected LBG (e.g., $b\sim1.7$ and 2.0 for the two ODIN samples), making a multi-tracer probe potentially possible at high-$z$. 

However, due to their faint continuum emission and the limited number of spectral features (often just the Ly$\alpha$ line), the selection of LAEs requires deep narrow-band imaging (e.g., \cite{Lee2023}) or IFS (e.g., \cite{Bacon2023}) observations. Most ongoing LAE surveys either cover much smaller areas than the MUST footprint (e.g., ODIN; \cite{Lee2023}) or are not deep enough for MUST (e.g., JPAS; \cite{Torralba-Torregrosa2023, Torralba-Torregrosa2024}). Furthermore, the bias of LAEs has only been measured on relatively small fields ($\lesssim100 {\rm deg}^2$, e.g., \cite{Ouchi2018, White2024}). We do not include LAEs in our current cosmological forecast for these reasons. At the same time, this implies that the cosmological constraining power of MUST could be further enhanced when a wide \& deep LAE sample becomes available in the future. We should note that the DECam on the 4 m Blanco telescope (e.g., \cite{DECam}) and the HSC on the 8.2 m Subaru (e.g., \cite{2018PASJ...70S...4A, Aihara2022}) can carry out deep narrow- or medium-band surveys in the following years. It is worth noting that DESI has conducted pilot spectroscopic observations of LAEs, demonstrating their promise as LSS tracers of the high-redshift Universe.

\subsubsection{Quasi-Stellar Objects (QSO)}
    \label{ssec:qso}

Quasi-stellar objects (QSOs) or quasars\footnote{We use the term QSO and quasar interchangeably in this work.} are both direct tracers of the dark matter field and sources for Lyman-$\alpha$ forest detections. The transition in their usage — from tracers to sources — typically occurs around $z \sim 2$, depending on their comoving densities. QSOs have been an essential ingredient in major spectroscopic surveys, serving as primary targets for probing the $z \gtrsim 2$ Universe. The Stage-III and IV spectroscopic surveys, such as BOSS \cite{ross12}, eBOSS \cite{myers15}, DESI \cite{yeche20, DESI_QSO}, and WEAVE \cite{weave_white_paper, weave-qso} define the modern standard of QSO target selection using multi-band deep imaging data and broad-band color cuts to isolate quasar candidates from stars and other galaxies (e.g., \cite{fan06, jiang08, willott10, carnell15, reed15, mat16, banados16, wang16, yang18}). At the same time, these surveys have explored QSO selection based on flux variability (e.g., \cite{Palanque-Delabrouille2016}) and the use of machine learning algorithms (e.g., \cite{yeche10, jin19, DESI_QSO}). 

While QSOs will certainly enable a wide range of interesting scientific topics in the age of Stage-V surveys, their low 3D density and color degeneracy with abundant Milky Way stars at $z>2$ (e.g., see Figure~\ref{fig:nz_bias}) make them less appealing LSS tracers. Therefore, we do not include QSO in the Fisher forecast to constrain cosmological parameters. In Figure~\ref{fig:nz_bias}, we adopt the pure luminosity function evolution model in \cite{Palanque-Delabrouille2016} to estimate the 3D density of QSOs at $r<23.5$. Assuming 80\% of the QSOs will be observed, MUST will reach a QSO density of $\sim$310 per deg$^{-2}$ with $\sim$90 deg$^{-2}$ at $z > 2.1$, including the QSOs already observed by BOSS/eBOSS and DESI ($\sim$60 deg$^{-2}$ at $z > 2.1$). 

It is worth noting that, while not being the most promising high-$z$ LSS tracers, at $z>2.1$, QSOs provide a vital capability for studying the IGM using the Ly$\alpha$ forest, which contains intriguing cosmological potential such as constraining the nature of dark matter. We will forecast the potential of Ly$\alpha$-QSO and provide a more thorough discussion in the following work of this series. 

\begin{table*}[!htb]
\centering
\resizebox{\textwidth}{!}{%
\begin{tabular}{|c|c|c|c|c|c|c|c|c|c|c|}
\hline
\rowcolor[HTML]{ECF4FF} 
\textbf{\begin{tabular}[c]{@{}c@{}}Sample \\ Name\end{tabular}} &
  \textbf{\begin{tabular}[c]{@{}c@{}}Magnitude Limit\\ (AB mag)\end{tabular}} &
  \textbf{\begin{tabular}[c]{@{}c@{}}Redshift\\ Distribution\end{tabular}} &
  \textbf{\begin{tabular}[c]{@{}c@{}}Angular Density\\ (deg$^{-2}$)\end{tabular}} &
  \textbf{\begin{tabular}[c]{@{}c@{}}Number of\\ Redshift Bins\end{tabular}} &
  \textbf{\begin{tabular}[c]{@{}c@{}}3D Density\\ ($10^{-3}$\ h$^{3}$ Mpc$^{-3}$)\end{tabular}} &
  \textbf{\begin{tabular}[c]{@{}c@{}}Bias \\ Value\end{tabular}} &
  \textbf{\begin{tabular}[c]{@{}c@{}}Survey Area\\ (deg$^2$)\end{tabular}} &
  \textbf{\begin{tabular}[c]{@{}c@{}}$T_{\rm Exp}$\\ (Hour)\end{tabular}} &
  \textbf{\begin{tabular}[c]{@{}c@{}}Total Number\\ of Redshift ($10^6$)\end{tabular}} &
  \textbf{\begin{tabular}[c]{@{}c@{}}Total Fiber\\ Hours ($10^6$)\end{tabular}} \\ \hline
\rowcolor[HTML]{FFCB2F} 
\textbf{BGS} &
  $20.18 < r < 21.0$ &
  $0.1<z<0.7$ &
  2,000 &
  2 &
  5.7 &
  1.6 &
  $\sim$13,000 &
  0.04 &
  26.0 &
  1.04 \\ \hline
\rowcolor[HTML]{FFCCC9} 
\textbf{LRG} &
  $21.6 < z_{\rm fib} < 22.8$ &
  $0.8<z<1.3$ &
  2,000 &
  2 &
  1.0 &
  2.8 &
  $\sim$13,000 &
  1.15 &
  26.0 &
  30.0 \\ \hline
\rowcolor[HTML]{99DAED} 
\cellcolor[HTML]{99DAED} &
  \cellcolor[HTML]{99DAED} &
  \cellcolor[HTML]{99DAED} &
  Opt:3,600 &
  2 &
  1.49 &
  1.5 &
  $\sim$13,000 &
  0.32 &
  46.8 &
  15.0 \\ \cline{4-11} 
\rowcolor[HTML]{99DAED} 
\multirow{-2}{*}{\cellcolor[HTML]{99DAED}\textbf{ELG}} &
  \multirow{-2}{*}{\cellcolor[HTML]{99DAED}$24.1 < g_{\rm fib} < 24.6$} &
  \multirow{-2}{*}{\cellcolor[HTML]{99DAED}$0.8<z<1.4$} &
  Con:2,500 &
  2 &
  1.03 &
  1.5 &
  $\sim$11,000 &
  0.30 &
  27.5 &
  8.25 \\ \hline
\rowcolor[HTML]{CBCEFB} 
\cellcolor[HTML]{CBCEFB} &
  \cellcolor[HTML]{CBCEFB} &
  \cellcolor[HTML]{CBCEFB} &
  Opt:1,500 &
  3 &
  0.21 &
  \cellcolor[HTML]{CBCEFB} &
  $\sim$11,000 &
  2.50 &
  15.6 &
  39.0 \\ \cline{4-6} \cline{8-11} 
\rowcolor[HTML]{CBCEFB} 
\multirow{-2}{*}{\cellcolor[HTML]{CBCEFB}\textbf{$u$-Dropout}} &
  \multirow{-2}{*}{\cellcolor[HTML]{CBCEFB}$22.4 < r < 24.5$} &
  \multirow{-2}{*}{\cellcolor[HTML]{CBCEFB}$2.2<z<3.5$} &
  Con:600 &
  3 &
  0.11 &
  \multirow{-2}{*}{\cellcolor[HTML]{CBCEFB}4.5} &
  $\sim$8,000 &
  2.50 &
  6.0 &
  15.0 \\ \hline
\rowcolor[HTML]{9AE881} 
\cellcolor[HTML]{9AE881} &
  \cellcolor[HTML]{9AE881} &
  \cellcolor[HTML]{9AE881} &
  Opt:870 &
  2 &
  0.16 &
  \cellcolor[HTML]{9AE881} &
  $\sim$11,000 &
  5.8 &
  10.4 &
  60.3 \\ \cline{4-6} \cline{8-11} 
\rowcolor[HTML]{9AE881} 
\multirow{-2}{*}{\cellcolor[HTML]{9AE881}\textbf{$g$-Dropout}} &
  \multirow{-2}{*}{\cellcolor[HTML]{9AE881}$23.0 < i < 24.6$} &
  \multirow{-2}{*}{\cellcolor[HTML]{9AE881}$3.2<z<4.5$} &
  Con:300 &
  2 &
  0.06 &
  \multirow{-2}{*}{\cellcolor[HTML]{9AE881}5.3} &
  $\sim$8,000 &
  5.80 &
  3.0 &
  17.4 \\ \hline
\rowcolor[HTML]{FFACAC} 
\cellcolor[HTML]{FFACAC} &
  \cellcolor[HTML]{FFACAC} &
  \cellcolor[HTML]{FFACAC} &
  Opt:390 &
  1 &
  0.06 &
  \cellcolor[HTML]{FFACAC} &
  $\sim$11,000 &
  5.8 &
  2.6 &
  15.1 \\ \cline{4-6} \cline{8-11} 
\rowcolor[HTML]{FFACAC} 
\multirow{-2}{*}{\cellcolor[HTML]{FFACAC}\textbf{$r$-Dropout}} &
  \multirow{-2}{*}{\cellcolor[HTML]{FFACAC}$23.0 < z < 25.0$} &
  \multirow{-2}{*}{\cellcolor[HTML]{FFACAC}$4.0<z<5.5$} &
  Con:100 &
  1 &
  0.03 &
  \multirow{-2}{*}{\cellcolor[HTML]{FFACAC}6.4} &
  $\sim$8,000 &
  5.80 &
  1.0 &
  5.8 \\ \hline
\rowcolor[HTML]{DAE8FC} 
QSO &
  $r<23.5$ &
  $2.0 < z < 5.0$ &
  90 &
   &
   &
   &
  $\sim$13,000 &
  0.32 &
  1.2 &
  0.38 \\ \hline
\end{tabular}%
}
\caption{Summary of the low- and high-redshift samples selected in this work. All but the QSO samples are considered in the cosmological forecast. For the ELG, $u$-, $g$, and $r$-dropout galaxies, we estimate an optimistic (``Opt'') and conservative (``Con'') surface density to design the survey.}
\label{table:sample}
\end{table*}

\subsection{Summary of the Target Selection}
    \label{ssec:targetsum}

In Table~\ref{table:sample}, we summarize the preliminary target selection discussed above and provide an initial estimate of the survey cost, measured in total fiber-hours. All but the QSO samples are considered in the cosmological forecast.
As the instrument design for MUST's focal plane and spectrograph system is still in the conceptual phase, a realistic exposure time calculator (ETC) based on mock spectra is under development. Here, we adopt a more empirical approach to estimate the required exposure time for MUST to secure the redshifts of the LSS tracers discussed above.

While MUST's effective collecting area is $\sim 2.6$ times of DESI\footnote{We assume the DESI primary mirror has a diameter of 3.8\,m with a central obscuration of 1.8\,m. For MUST, we adopt a central obscuration of 2.6\,m.}, many other systematic factors further complicate the scaling of exposure time. Using DESI's mock spectra framework \texttt{specsim}\footnote{\url{https://github.com/desihub/specsim}} with MUST's optical specifications, fiber diameters, and Lenghu's typical seeings, we compare the median per-pixel signal-to-noise ratios (SNRs) in the three channels between DESI and MUST for different ELGs, LRGs, and QSOs by assuming their spectrographs share the same performance. For the same exposure time, MUST typically shows $2.2\times$ to $2.4\times$ of improvement in SNR. The current \texttt{specsim} setup is still highly oversimplified. Lenghu should have less light pollution than Kitt Peak, hence its typical sky brightness should be an advantage for MUST. Lenghu's high-altitude location will also reduce atmospheric attenuation at the blue end, which could be critical for the high-redshift LSS survey that relies on the SNR of the Ly$\alpha$ emission lines. Meanwhile, DESI's CCD detector for the NIR channel still has a significantly higher quantum efficiency (QE) at $\lambda > 850$\,nm. Moreover, the median SNR does not straightforwardly translate into improvement in redshift precision or success rate.

To be on the conservative side, we simply assume that MUST can reach the same SNR 2.2 times faster than DESI, and the exposure time will increase $10^{2 \delta{m}/2.5}$ times for the fainter sample, where $\delta{m}$ is the magnitude difference. For the BGS sample, DESI achieves a $\>95$\% redshift success rate with a 200s effective exposure time ($T_{\rm eff}$) for the $r<20.175$\,mag sample. Therefore, MUST will require a $\sim 145$s ($\sim 1030$s) for the $r<21.0$\,mag ($r<21.5$\,mag) BGS sample. Similarly, assuming MUST's LRG, ELG, and QSO samples are $1.2$\,mag, $0.5$\,mag, and $0.5$\,mag fainter than DESI's selection limits, DESI's 1000s nominal $T_{\rm eff}$ will translate into $\sim 4150$s, $\sim 1140$s, and $\sim 4150$s for MUST.


Estimating the exposure time for the LBGs (or LAEs) is more difficult. Based on the recent LBG observation campaign by DESI \cite{Ruhlmann-Kleider2024}, for a $r<24.2$\,mag $u$-dropout sample, DESI can already achieve a $\sim 70$\% redshift efficiency in 2.5 hours of effective exposure time. However, the efficiency seems to plateau after 2.5 hours, as a 5-hour $T_{\rm eff}$ still cannot improve it to $>80$\%. It is worth noting that there is certainly room to improve the LBG redshift estimation in the near future with the arrival of more accurate spectral models or templates. Therefore, for the $u$-dropout sample, we apply a more conservative empirical scaling relation and assume that MUST can achieve the same success rate for our $r < 24.5$\,mag sample with a 1.97-hour $T_{\rm eff}$. For the $g$- and $r$-dropouts, it is less straightforward to estimate the required exposure time as there is no statistically meaningful DESI sample to compare with. In \cite{Wilson2019}, the authors estimated that an 80-minute exposure would be enough for a $z=24.0$\,mag galaxy using a 6.5\,m telescope. Scaling this up to a $z<25.0$\,mag sample would require an effective exposure time of $\sim3.8$ hours. However, given the systematic uncertainties involved in LBG observations and redshift measurements, we apply a 30\% margin to their expected exposure times, making it a 2.5-hour $T_{\rm eff}$ for MUST's $u$-dropouts and 5.8-hour $T_{\rm eff}$ for the $g$- and $r$-dropouts.

\subsection{Conceptual Survey Design}
    \label{ssec:surveydesign}

\begin{figure*}[!htb]
    \includegraphics{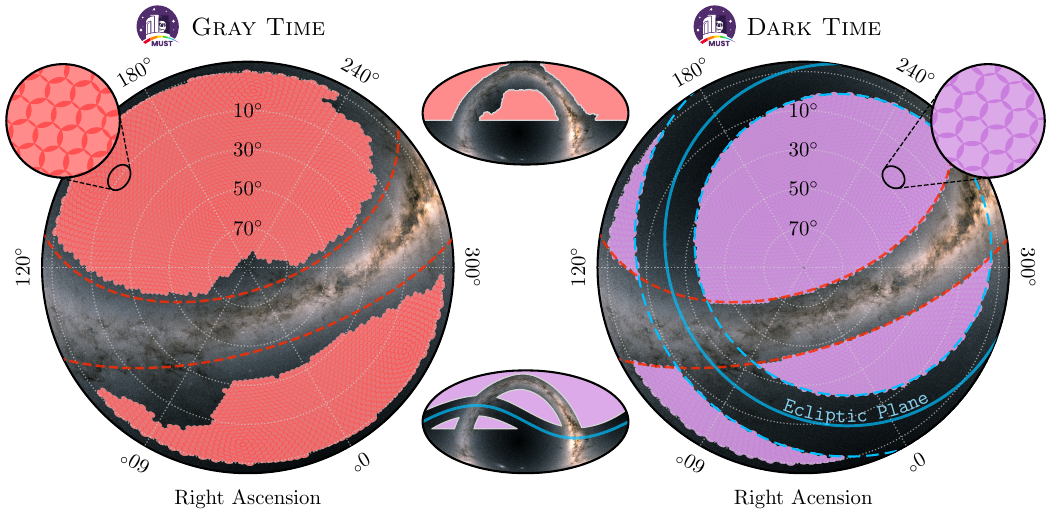}
    \caption{Footprint and an example tiling for MUST gray-time ({\it left}) and dark-time ({\it right}) observations. Each filled circle represents the sky projection of the MUST focal plane (5\,deg$^2$). The gray-time footprint is based on the DECaLS imaging sample, while the dark-time footprint follows the expected CSST sky coverage. Both footprints are restricted to declinations above $-10^\circ$, reflecting the latitude constraint of the Lenghu site. Background color maps indicate the {\it Gaia} EDR3 stellar density \cite{Gaia_EDR3}. The gray-time and dark-time tilings yield around 3,500 and 3,000 pointings, covering $\sim$13,000 and 11,000\,deg$^{2}$, respectively.}
    \label{fig:MUST_Survey_design}
\end{figure*}

This section briefly describes the conceptual survey designs for the bright-gray and dark-time programs of MUST based on the previous description of the telescope \& instrument design and the target selection under several oversimplified assumptions. In the upcoming work in this series, we will present a more robust survey design, backed by detailed instrument specifications, an exposure time calculator (ETC) for MUST, and an improved target selection strategy. 

\subsubsection{Dark Time Survey}
    \label{ssec:darktime}

We assume that the CSST photometry will support the selection of dark-time targets for MUST, which will mainly include the LBG at $2.2<z<5.0$. The right panel of Figure~\ref{fig:MUST_Survey_design} shows a preliminary observing pattern, where each pointing corresponds to the 5\,deg$^2$ effective FoV of MUST.
Given the Lenghu site's latitude, MUST cannot access the full CSST footprint. Assuming a maximum airmass of 1.5 ($\sim$40$^{\circ}$ above the horizon) for observation, we restrict the footprint to declinations above $-10^{\circ}$. Surveying 11,000\,deg$^{2}$, or 3,000 pointings with a minimal overlap for a continuous coverage, and adopting an average exposure time of 2.5 hours per target, this requires a total of 750 effective dark nights, with an average of 10 observing hours per night. Assuming a 70\% clear-night fraction and 35\% dark-time availability per year that accounts for an 80\% survey efficiency, this corresponds to an 8.5-year observing program. In total, this dark-time survey is expected to produce $\sim 20$ million redshift measurements under the more optimistic high-redshift target-selection scenario, assuming an overall redshift-measurement efficiency of 70\%.

\subsubsection{Gray Time Survey}
    \label{graytime}

As the gray-time program will primarily focus on BGS, LRG, and ELG, we use the Legacy Survey footprint to simulate the pointings needed for MUST, as shown in the left panel of Figure~\ref{fig:MUST_Survey_design}. With the declination limited to $> -10^\circ$, MUST can access a total of $\sim$13,000\,deg$^{2}$ of the Legacy Survey footprint, corresponding to $\sim$3,500 contiguous MUST pointings.
With an average exposure time of 1 hour per target and a target density of approximately twice the fiber density of MUST ($\sim$4,000 deg$^{-2}$), which may include, for example, 2,000\,deg$^{-2}$ BGS, 2,000\,deg$^{-2}$ LRG, 3,600\,deg$^{-2}$ ELG, and 400\,deg$^{-2}$ of other targets, the program would require a total of 700 effective gray nights with 10 observing hours per night. Assuming 40\% of gray time per year, that accounts for an overall 80\% survey efficiency and a 70\% clear-night fraction, the gray-time survey would span $\sim$7 years. In total, this program is expected to accumulate $\sim$88 million redshift measurements, assuming an overall 90\% redshift measurement efficiency.

Among the gray-time targets, LRG requires significantly longer exposure time (e.g., $\sim$1 hour for a target with $z_{\rm fiber}=22.8$\,mag) compared to the others. With a carefully optimized survey strategy, the average exposure time per gray-time target could be reduced, speeding up the current program and leaving room for a further increase in target density. For example, within the current detection limits of the Legacy Survey, we could select BGS down to $r<21.5$\,mag with a density of $\sim$4,000\,deg$^{-2}$ and LRG down to $z_{\rm fiber}=23.0$\,mag with a density of $\sim$3,000\,deg$^{-2}$. Combining them with the current ELG sample yields a total target density $\sim$2.7 times the fiber density. Assuming a 1-hour \emph{average} exposure time, this enhanced program will require $\sim$9.5 years of gray time to finish and yield a $\sim$114 million redshifts at $z<1.6$ with an 80\% average redshift efficiency. Safely assuming a fraction of these gray-time targets can be observed during bright time, it is possible to complete this program alongside the 8.5-year dark-time survey.

Putting the gray and dark programs together, MUST has the capability of collecting $\sim$110 million redshifts at $0.1 < z < 5.0$ before 2040 (see Figure~\ref{fig:surveys}, with a comparison with historical and upcoming spectroscopic surveys).

\section{Cosmological Forecasts}
    \label{sec:forecast}

To quantitatively evaluate the scientific potential of MUST in addressing the fundamental cosmological questions outlined in Section~\ref{sec:motivation}, we forecast statistical errors for several key science cases based on the expected target densities and biases detailed in Section~\ref{sec:target}, with a Planck 2018 $\Lambda$CDM fiducial cosmology.\footnote{Marginalized means of parameters from the \texttt{Plik} likelihood, see Table 1 of \cite{Planck2020c}.}
We produce two sets of forecasts — optimistic and conservative — based on different estimates of the available imaging data to ensure a comprehensive evaluation of MUST's scientific potential.
To illustrate the potential improvements achievable by MUST, we compare our forecasts with those for DESI, the current state-of-the-art Stage-IV survey, using its target specifications from \cite{DESIcoll2016}. Specifically, we focus on the DESI galaxy samples, i.e., BGS, LRG, and ELG.

To fully assess the constraining power of next-generation cosmological data, we incorporate external CMB priors from Planck \cite{planck2020d} and Simons Observatory (SO; \cite{SO2018}) into our Fisher forecasts, following the methodology in \cite{Sailer+2021}.
Additionally, we perform Bayesian cosmological parameter inferences using Markov-Chain Monte-Carlo (MCMC) methods without these CMB priors, and directly compare our forecasts to constraints derived from the existing Planck and Atacama Cosmology Telescope (ACT; \cite{ACT2024}) CMB data, as well as the Pantheon+ \cite{Pantheon+2022} Type Ia SN compilation.

\subsection{Methodology}
    \label{ssec:forecastmethod}

\subsubsection{Fisher forecast}

The Fisher forecast method is widely used to estimate the statistical uncertainty of cosmological parameters inferred from expected clustering measurements. This method is based on the Cram\'{e}r-Rao inequality \cite{Rao1945, Cramer1946}, which establishes that no unbiased estimator can achieve a covariance matrix smaller than the inverse of the Fisher information matrix $\mathbf{F}$. Consequently, the standard deviation of the $i$-th measured parameter, $\theta_i$, is bounded from below by $\sigma(\theta_i) \ge \sqrt{(F^{-1})_{ii}}$. Following the formulation in \cite{Tegmark1997}, the Fisher information matrix for a redshift survey can be approximated as
\begin{equation}
    F_{ij} \sim \frac{\partial\boldsymbol{P}^\mathrm{T}}{\partial\theta_i}\mathbf{C}^{-1}\frac{\partial\boldsymbol{P}}{\partial\theta_j},
\end{equation}
where $\boldsymbol{P}$ is the data vector with $P_i = P(k_i, \mu_i)$ representing the power spectrum of the $i$-th bin in $\boldsymbol{k}$-space, and $\mathbf{C}$ is the corresponding covariance matrix of the power spectrum. With the usual assumption that the overdensity field is approximately a Gaussian random field, the covariance matrix $\mathbf{C}$ is given by

\begin{equation}
    C_{ij} \sim \delta_{ij}\frac{2P_i^2}{V_iV_{\rm eff}},
    \label{eq:covariance}
\end{equation}
where $V_i$ is the volume of the $i$-th bin in $\boldsymbol{k}$-space, and $V_{\rm eff}$ is the effective survey volume.

Throughout this work, we perform Fisher forecasts using the tracer power spectrum $P(\boldsymbol{k})$ with 
\begin{equation}
    k_{\rm max} = 0.4\,h\,{\rm Mpc}^{-1},
\end{equation}

which corresponds to the smallest scale where a state-of-the-art perturbation theory achieves sub-percent accuracy (e.g. \cite{EFT2019}).
In our forecasts, the tracer power spectrum is computed using either full-shape models or a template-based approach.
For the full-shape model, we use the forecast tool {\sc FishLSS}\footnote{\url{https://github.com/NoahSailer/FishLSS}} described in \cite{Sailer+2021}. {\sc FishLSS} computes the nonlinear tracer power spectrum using Lagrangian perturbation theory with a linear tracer bias model, refined by a single-parameter counter-term for small-scale corrections \cite{velocileptors_2020} that is calibrated to the HaloFit prescription \cite{halofit2012}. Shot noise and the finger-of-god effect are incorporated based on the density of the tracers.
The template-based approach is used only for the $f\sigma_8$ forecasts (Section~\ref{ssec:forecastgravity}). In this case, we use a code\footnote{\url{https://github.com/wdoumerg/Forecast\_highz\_spectroscopic\_survey}} developed for a recent spectroscopic survey forecast \cite{dAssignies+2023}, which refers to the RSD template from \cite{White-Song&Percival2009}.

\subsubsection{Cosmological inference}

To convert the cosmological distances from BAO and structure growth parameters from RSD into cosmological parameters, we use the cosmological inference code \textsc{Cobaya} \cite{Cobaya2021} with our forecast results, along with the Planck 2018 PR3 likelihoods\footnote{We use likelihoods from the temperature (TT), polarization (EE), and their cross-correlation (TE) power spectra for high-$\ell$, and TT and EE for low-$\ell$.} \cite{Planck2020l}, their combination with the ACT DR6 results with CMB lensing\footnote{We use the official `\texttt{actplanck\_baseline}' likelihood when including ACT results, see \url{https://github.com/ACTCollaboration/act\_dr6\_lenslike}.} \cite{ACT2024}, and the Pantheon+ SN data \cite{Pantheon+2022}.
For most cases, we use the Boltzmann code \textsc{camb} \cite{Lewis:1999bs} to evaluate the theoretical power spectra, while for modified gravity, we employ \textsc{ISiTGR} \cite{Dossett2011}. Bayesian inference is performed using the MCMC Metropolis sampler \cite{Lewis2013} in \textsc{Cobaya}, and the resulting chains are visualized as parameter posteriors with the \textsc{getdist} package \cite{Lewis2019}.

For the warm dark matter mass ($m_{\rm X}$) forecast, we implement the $m_{\rm X}$-dependent Ly$\alpha$ power spectrum and its covariance matrix, as described in Section~\ref{ssec:forecastwdm}, and use the \textsc{emcee} \cite{2013PASP..125..306F} toolkit for MCMC sampling.

\subsection{Cosmic Expansion History \& Dark Energy}
    \label{ssec:forecastbao}

\begin{figure*}[!htb]
    \centering
    \includegraphics{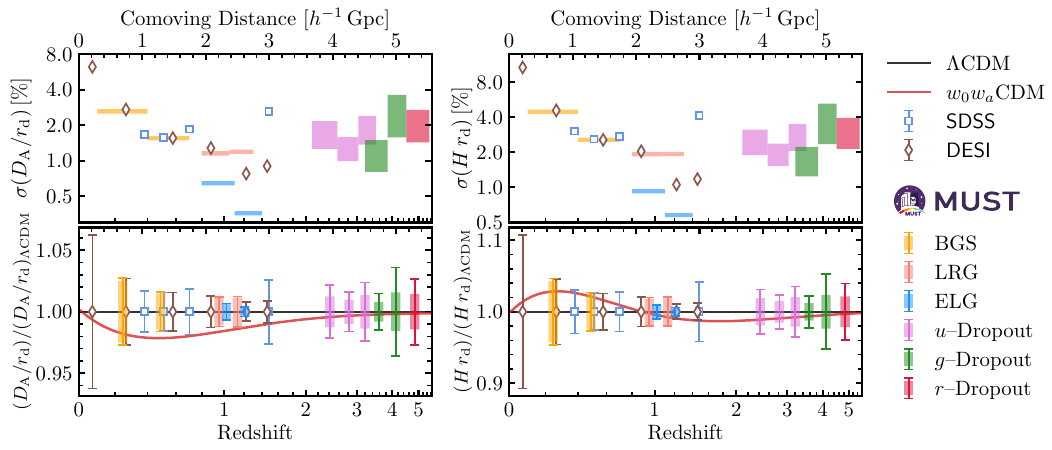}
    \caption{{\it Top} panels: expected uncertainties in $D_{\rm A} (z) / r_{\rm d}$ ({\it left}) and $H(z) r_{\rm d}$ ({\it right}) from DESI and MUST galaxy samples, compared with existing SDSS measurements \cite{eBOSS2021}. Shaded regions indicate the range between optimistic and conservative MUST forecasts. {\it Bottom} panels: forecasted distance measurements assuming the fiducial $\Lambda$CDM cosmology, compared with predictions from the best-fit $w_0 w_a$CDM model obtained in the DESI Y1 analysis \cite{DESI2024c}. For MUST, error bars indicate conservative forecasts, while boxes denote optimistic results.}
    \label{fig:DA_Hz_measure}
\end{figure*}

\begin{figure*}[!htb]
    \centering
    \includegraphics{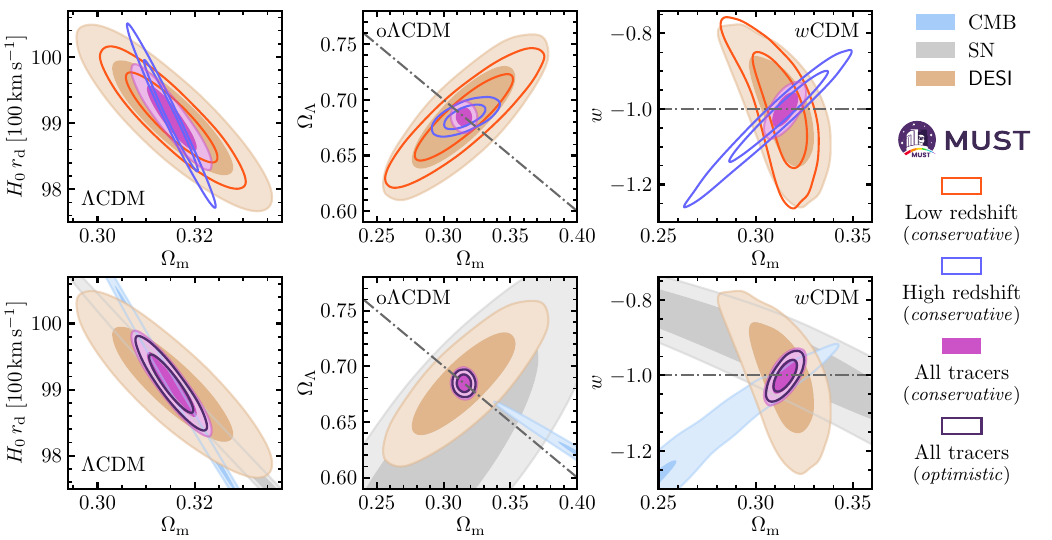}
    \caption{Cosmological parameter constraints from our DESI and MUST distance forecasts. All distance forecasts assume the fiducial $\Lambda$CDM cosmology, while parameter inferences are performed in three models: $\Lambda$CDM ({\it left}), o$\Lambda$CDM ({\it center}), and $w$CDM ({\it right}). {\it Top} panels: constraints from MUST low-redshift ($z \lesssim 1.5$) and high-redshift ($z \gtrsim 2$) galaxy samples are shown separately to highlight the contribution of the novel high-redshift tracers. {\it Bottom} panels: comparisons of our forecasts with current constraints from Planck 2018 PR3 \cite{Planck2020l} and Pantheon+ \cite{Pantheon+2022} data. Gray dash-dotted lines indicate the $\Lambda$CDM parameter space.}
    \label{fig:cospar_BAO}
\end{figure*}

As a cosmic ``standard ruler'' (see Section~\ref{ssec:darkenergy}), the BAO feature provides geometric measurements at various redshifts, revealing the cosmic expansion history and constraining the density and evolution of cosmic components, particularly dark energy. The characteristic BAO scale corresponds to the sound horizon at the drag epoch, $r_{\rm d}$, and relates directly to the angular ($\Delta\theta$) and redshift ($\Delta z$) separations of tracer pairs that exhibit excess correlations in the transverse and line-of-sight directions, respectively. In a flat Universe, these relationships are given by
\begin{align}
& r_{\rm d}=(1+z) D_{\rm A} (z) \Delta\theta= \Delta\theta \int_0^z \frac{c\,{\rm d}z'}{H(z')}, \\
& r_{\rm d}\approx\frac{c \Delta z}{H(z)},
\end{align}
where $D_{\rm A}(z)$ is the angular diameter distance and $H(z)$ is the Hubble parameter. Thus, BAO measurements constrain the combinations $D_{\rm A} (z) /r_{\rm d}$ and $H(z) r_{\rm d}$ as functions of redshift. 

Figure~\ref{fig:DA_Hz_measure} presents the forecasted uncertainties in $D_{\rm A} (z)/r_{\rm d}$ and $H(z)r_{\rm d}$ for different MUST and DESI tracers, compared with existing SDSS measurements \cite{eBOSS2021}. In the nearby Universe ($z \lesssim 0.7$), the precision from MUST is comparable to that of current surveys, primarily due to limitations imposed by cosmic variance. At intermediate redshifts ($0.7 \lesssim z \lesssim 1.5$), the high-density ELG samples from MUST are expected to provide a twofold improvement in distance measurement precision. Importantly, MUST will deliver unprecedented, percent-level measurements of the cosmic expansion history over $2 \lesssim z \lesssim 5.5$, offering crucial geometric constraints spanning from cosmic noon to the post-reionization era.
Notably, it is at high redshift where the difference between our optimistic and conservative forecasts becomes most significant -- nearly a factor of two. This reflects the challenges and uncertainties associated with imaging surveys of high-redshift galaxies and quasars.

The impact of these distance measurements on cosmological parameter constraints is shown in Figure~\ref{fig:cospar_BAO}, for the $\Lambda$CDM cosmology and its two one-parameter extensions: o$\Lambda$CDM and $w$CDM, which include $\Omega_k$ and $w$ as additional parameters, respectively. In all cases, significant improvements over DESI forecasts are observed. These gains are primarily driven by the inclusion of high-redshift samples, which alter the degeneracy directions due to the redshift dependence of parameter constraints. As a result, the statistical errors of $\Omega_k$ and $w$ are reduced by approximately 88\% and 70\%, respectively, when comparing MUST to DESI. The optimistic and conservative forecasts for MUST yield nearly identical results.

\begin{figure}[H]
    \centering
    \includegraphics{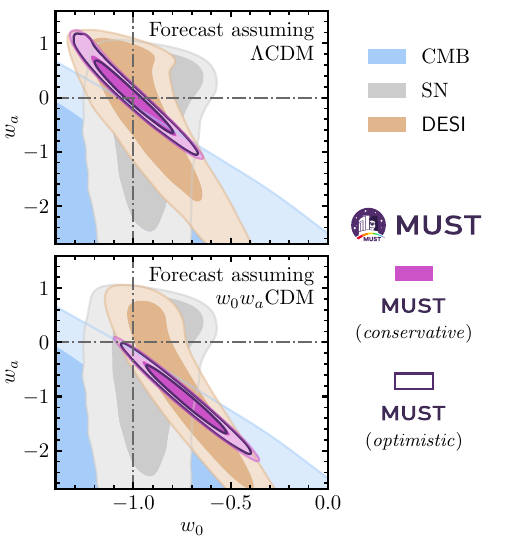}
    \caption{Constraints on $w_0$ and $w_a$ from parameter inferences in the $w_0 w_a$CDM model, based on our DESI and MUST distance forecasts, compared with existing Planck 2018 PR3 \cite{Planck2020l} and Pantheon+ \cite{Pantheon+2022} results. Forecasts are performed assuming either the fiducial $\Lambda$CDM cosmology ({\it top}) or the best-fit $w_0 w_a$CDM model from DESI Y1 data \cite{DESI2024c} ({\it bottom}). Gray dash-dotted lines indicate the $\Lambda$CDM parameter values.}
    \label{fig:cospar_DE}
\end{figure}

Note that we do not attempt a full joint constraint from the combination of CMB, SN, and BAO results in this work. As shown in Figure~\ref{fig:cospar_BAO}, the CMB and SN posteriors already show mild tension (beyond $1\sigma$) under simple extensions to $\Lambda$CDM. In such cases, Fisher combinations depend largely on the chosen fiducial model. If we were to add the MUST BAO forecasts to the combination, the visual impression of the joint contours would be sensitive to where the fiducial is set. Since MUST improves BAO precision substantially relative to DESI, any existing mild tension could become more apparent in a joint display.
For this reason we use Planck 2018 $\Lambda$CDM as the baseline and focus on single-probe BAO forecasts to quantify the information gain from MUST. We treat the bottom panels in Figure~\ref{fig:cospar_BAO} as illustrative and note their sensitivity to the fiducial choice. There is currently no consensus on the central value for o$\Lambda$CDM or $w$CDM across different probes. If future data favor an offset from our baseline, the tensions may be more prominent and point towards unaccounted systematics or new physics. In contrast, if the posteriors converge, the MUST precision will translate into tighter combined constraints. 

To evaluate the potential contribution of MUST to the recent indications of evolving dark energy reported by DESI \cite{DESI2024c, DESI2025}, we perform cosmological inferences in the flat-$w_0w_a$CDM cosmology under two scenarios. In the first, we use our fiducial $\Lambda$CDM cosmology for the distance forecasts; whereas in the second, we assume the best-fit $w_0w_a$CDM cosmology from \cite{DESI2024c}. The results, shown in Figure~\ref{fig:cospar_DE}, are compared with DESI forecasts and current constraints from CMB and SN data. In both cases, the statistical precision of $w_0$ and $w_a$ is significantly improved relative to DESI, corresponding to a 6--7 times increase of the $w_0$--$w_a$ figure of merit. If the DESI best-fit $w_0w_a$CDM model reflects the true cosmology, MUST alone would achieve a $3.1\,\sigma$ deviation from $\Lambda$CDM in the optimistic case, and $2.7\,\sigma$ in the conservative case.

In summary, MUST will provide the most stringent constraints on evolving dark energy from purely geometrical measurements, significantly outperforming current CMB and SN results. It is worth noting, however, that the commonly used CPL parameterization in Eq.~\eqref{eq:de_cpl} may bias measurements of the high-redshift behavior of the dark energy equation of state \cite{Linden2008}. Meanwhile, the time-evolving term $w_a$ enters as a factor of $(1 - a)$, which evolves slowly at high redshift, thereby limiting sensitivity to late-time changes (see the bottom panel of Figure~\ref{fig:DA_Hz_measure}). Since the primary advantage of MUST lies in its high-redshift ($z \gtrsim 2$) coverage, it is important to explore alternative parameterizations of $w(z)$ that better capture its high-redshift evolution. We leave such investigations to future work.

\subsection{Structure Growth \& Modified Gravity}
    \label{ssec:forecastgravity}

\begin{figure*}[!htb]
    \centering
    \includegraphics{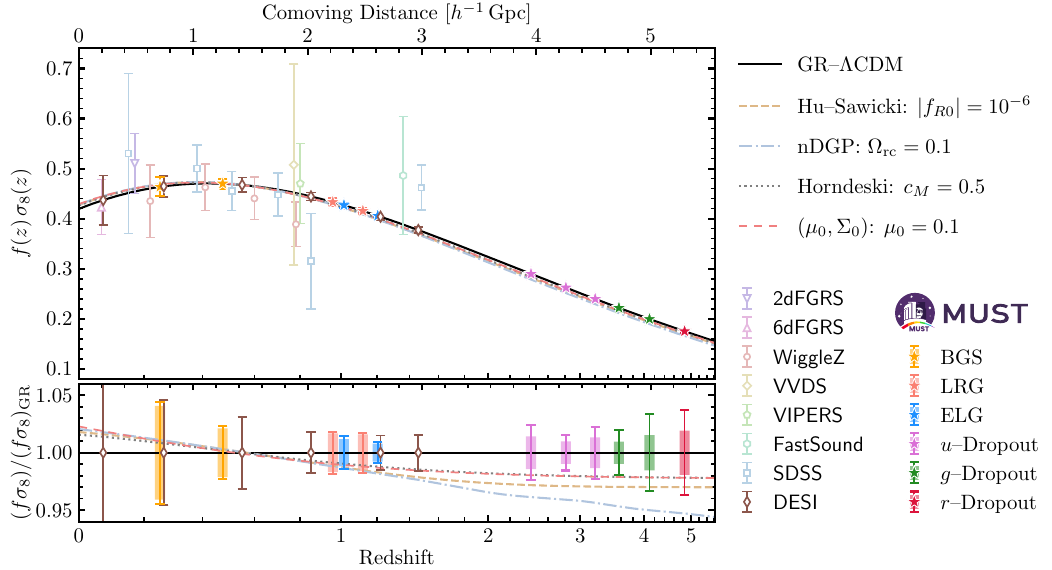}
    \caption{Fisher forecasts for $f(z) \sigma_8 (z)$ measurements from MUST and DESI galaxy samples. {\it Top} panel: comparison with existing measurements from SDSS \cite{eBOSS2021, SDSS16_fs8_1, SDSS16_fs8_2}, 2dFGRS \cite{2dFGRS_fs8, RSD_revive_2009}, VVDS \cite{VVDS_fs8}, VIPERS \cite{VIPERS_fs8}, WiggleZ \cite{WiggleZ_fs8}, 6dFGRS \cite{6dFGS_fs8}, and FastSound \cite{fsigma8_modified_gravities_2016}. {\it Bottom} panel: forecasted relative errors on $f(z) \sigma_8 (z)$ compared to the fiducial GR-$\Lambda$CDM cosmology. Error bars and shaded boxes represent conservative and optimistic forecasts, respectively.
    Also shown are predictions from various modified-gravity models that are broadly consistent with current cosmological observations, including Hu--Sawicki $f(R)$ gravity, nDGP gravity, the no-braiding running-Planck-mass Horndeski model in the $\alpha$-basis with $\alpha_B=\alpha_T=0$ and $c_K=10^{-2}$, and the phenomenological $(\mu_0,\Sigma_0)$ parametrization with $\Sigma_0=0$, all assuming the same $\Lambda$CDM background. The theoretical curves are computed using the \textsc{MGCAMB} \cite{MGCAMB2023} and $\mathcal{H}$-\textsc{EFTCAMB} \cite{EFTCAMB2026} packages, and are renormalized to current measurements to emphasize redshift evolution rather than absolute amplitude.}
    \label{fig:fsigma8_forecast}
\end{figure*}

Differences between GR and alternative gravity models are often reflected in their predicted growth histories. In practice, anisotropic galaxy clustering provides redshift-space-distortion constraints that are commonly summarized as measurements of $f\sigma_8(z)$. Figure~\ref{fig:fsigma8_forecast} shows Fisher forecasts for the statistical uncertainties in $f\sigma_8(z)$ from different MUST and DESI galaxy samples. The results are compared with existing constraints from previous surveys and with predictions from several representative modified-gravity scenarios computed on a $\Lambda$CDM background, including Hu--Sawicki $f(R)$ gravity for chameleon screening, normal-branch DGP (nDGP) gravity for Vainshtein screening, a running-Planck-mass Horndeski model, and a phenomenological $(\mu_0,\Sigma_0)$ parametrization with $\Sigma_0=0$.
For the parameter choices shown in Figure~\ref{fig:fsigma8_forecast}, the nDGP, Horndeski, and phenomenological $\mu$ models remain broadly compatible with current cosmological constraints. Recent constraints include $\Omega_{\rm rc}\lesssim 0.2$ \cite{Piga2023}, $c_M \lesssim 1.1$ at 95\% C.L., and $\mu_0 = 0.05 \pm 0.22$ \cite{DESI2024MG}. For Hu--Sawicki $f(R)$ gravity, current cosmological constraints reach $|f_{\rm R0}| \lesssim 5 \times 10^{-6}$ at 95\% C.L. \cite{Vogt_fR}, while astrophysical screening tests can be substantially stronger \cite{Desmond2020}. We therefore keep $f(R)$ primarily as a standard benchmark model.

We observe trends similar to those seen in the distance forecasts discussed in Section~\ref{ssec:forecastbao}. Specifically, the performance of MUST is comparable to that of DESI at $z \lesssim 0.7$. There are modest improvements at $0.7 \lesssim z \lesssim 1.5$, which could be further enhanced through future multi-tracer analyses combining different galaxy populations from both MUST and DESI. The most significant contribution from MUST lies in its ability to achieve percent-level structure growth measurements at high redshift ($z \gtrsim 2$), a regime that remains largely unexplored.
As with the distance forecasts, uncertainties in the availability of high-redshift imaging data lead to substantial differences between optimistic and conservative error estimates. Nonetheless, these additional high-redshift measurements will be essential for distinguishing among modified gravity models that remain consistent with current observations.

\begin{figure}[H]
    \centering
    \includegraphics{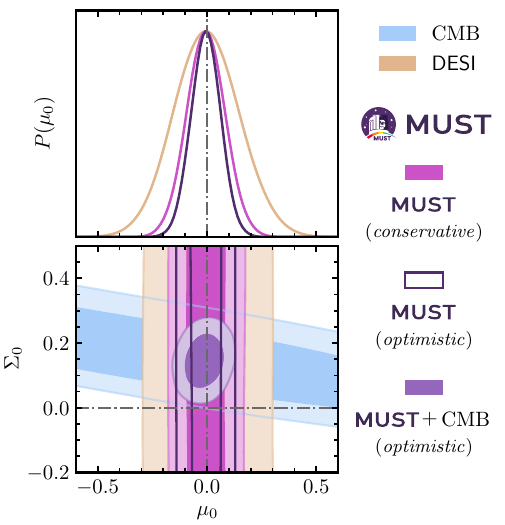}
    \caption{Constraints on $\Sigma_0$ and $\mu_0$ derived from BAO and RSD forecasts for MUST and DESI, assuming the fiducial $\Lambda$CDM cosmology. Results are compared with current CMB constraints from Planck PR3 \cite{Planck2020l} and ACT DR6 \cite{ACT2024}. The joint constraint from optimistic MUST forecasts combined with CMB data is also shown. Gray dash-dotted lines indicate the $\Lambda$CDM parameter values.}
    \label{fig:cospar_MG}
\end{figure}

We translate the forecasts of cosmological distances from BAO and of structure growth parameters from RSD into constraints on modified gravity using the $\mu(a, k)$ and $\Sigma(a, k)$ parameterization defined in Eqs.~\eqref{eq:mg_mu} and \eqref{eq:mg_sigma}. For simplicity, we assume that the distance and structure growth measurements are independent, and focus on a scale-independent form of $\mu$ and $\Sigma$, with the time dependence modeled as (e.g., \cite{Simpson2013}):
\begin{align}
\mu(a) &= \mu_0 \frac{\Omega_\Lambda(a)}{\Omega_\Lambda}, \label{eq:mg_mu0} \\
\Sigma(a) &= \Sigma_0 \frac{\Omega_\Lambda(a)}{\Omega_\Lambda}. \label{eq:mg_sigma0}
\end{align}
Constraints on $\mu_0$ and $\Sigma_0$ derived from our $f\sigma_8$ forecasts are shown in Figure~\ref{fig:cospar_MG}, along with existing CMB constraints from Planck 2018 DR3 \cite{Planck2020l} and ACT DR6 \cite{ACT2024}. As expected, galaxy clustering data constrain only the $\mu_0$ parameter. Compared to DESI, the expected uncertainty in $\mu_0$ is reduced by 41\% and 55\% under the conservative and optimistic MUST forecasts, respectively. When combined with CMB data, the improvements become 20\% and 30\%. These results highlight the strong potential of MUST for precision tests of General Relativity.

It is worth noting that, under the $\mu_0$ and $\Sigma_0$ parameterization adopted here, deviations from GR may appear less significant if the actual time evolution differs from the form assumed in Eqs.~\eqref{eq:mg_mu0} and \eqref{eq:mg_sigma0} \cite{Simpson2013}. To fully exploit the high-redshift structure growth measurements enabled by MUST, it will be important to explore alternative phenomenological parameterizations and physically motivated models of modified gravity in future work. For instance, as shown in Figure~\ref{fig:fsigma8_forecast}, MUST is expected to constrain the strength of $f(R)$ gravity, $|f_{R0}|$, down to the $10^{-6}$ level.

\subsection{Primordial Non-Gaussianity}
    \label{ssec:forecastfnl}

The presence of local-type PNG (see Section~\ref{ssec:primordial}) introduces a dependence on the initial gravitational potential field, $\phi$, into the galaxy bias model \cite{desjacques2018rvbias}. At leading order, this results in a so-called ``scale-dependent bias'' term in the galaxy power spectrum:
\begin{equation}
    P_{\rm gg}(k,z) = {\left(b_1+f_{\text{NL}}^{\text{local}} b_\phi \frac{3H_0^2 \Omega_{\rm m}}{2k^2 T(k) D(z)}\right)}^2 P_{\rm mm}(k,z) + P_{\rm shot}(z),
\end{equation}
where $b_1(z)$ denotes the linear galaxy bias, $b_\phi(z)$ quantifies the response of galaxy bias to $\phi$, $T(k)$ is the matter transfer function and $D(z)$ is the linear growth factor normalized to redshift $z$ during the matter-dominated epoch. The transfer function $T(k)$ is normalized to unity on large scales, where the PNG signal is most prominent, hence the scale-dependent bias typically follows a $k^{-2}$ dependence.
This effect can be readily probed through 2-point galaxy clustering statistics. For this reason, our forecasts focus exclusively on the local-shape PNG amplitude $f_{\rm NL}^{\rm local}$. For the scale-dependent bias term, we adopt the commonly used universality relation $b_\phi = 2\delta_{\rm c} (b_1-1)$, where $\delta_{\rm c}$ is the critical overdensity for spherical collapse. It is worth noting that recent studies suggest this relation may vary depending on the specific galaxy sample \cite{lazeyras2023bphi, adame2024unitpng}. We leave a detailed investigation of such effects for future work.

\begin{figure}[H]
    \centering
    \includegraphics{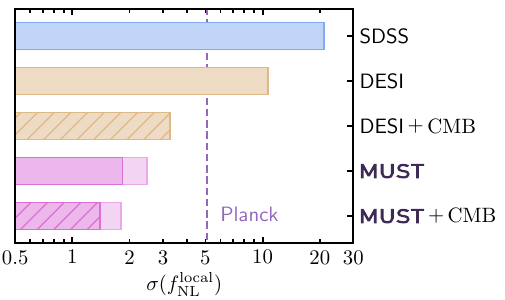}
    \caption{Forecasted $1\,\sigma$ constraints on $f_{\rm NL}^{\rm local}$ from MUST and DESI using galaxy power spectra, along with their combinations with CMB constraints from Planck low-$\ell$ \cite{planck2020d} and Simons Observatory (SO; \cite{SO2018}) high-$\ell$ measurements. Current constraints from SDSS \cite{Mueller2022} and Planck CMB data \cite{planck2020png} are also shown for comparison.}
    \label{fig:forecast_fNL}
\end{figure}

Our forecast results for MUST and DESI based on the galaxy power spectra are shown in Figure~\ref{fig:forecast_fNL}, alongside existing constraints from SDSS \cite{Mueller2022} and Planck \cite{planck2020png}. To assess the full potential of future PNG constraints, we further perform Fisher forecasts incorporating priors from Planck and Simons Observatory (SO; \cite{SO2018}) CMB observations, enabling joint constraints from galaxy spectroscopic and CMB data.
We find that the expected $1\sigma$ uncertainty on $f_{\rm NL}^{\rm local}$ from MUST alone ranges from $\sim$1.8 (optimistic) to $\sim$2.5 (conservative), depending on the assumed availability of imaging survey data. These values represent an approximate fivefold improvement over DESI and surpass the current CMB constraint of $\sigma(f_{\text{NL}}^{\text{local}}) = 5.1$ \cite{planck2020png}. When combined with CMB data, the constraint is expected to tighten further to $\sigma(f_{\text{NL}}^{\text{local}}) \sim 1.4$ to 1.8.
Additional methods not included in this forecast, such as the multi-tracer technique \cite{multitracer-PNG_2023} and bispectrum analyses \cite{fNL_bispec}, are expected to further enhance the constraining power of MUST. These approaches may allow $\sigma(f_{\text{NL}}^{\text{local}})$ to reach below unity, a critical threshold from a theoretical perspective, as a non-detection at this level would exclude many scenarios involving the curvaton mechanism \cite{kawasaki2011curvaton, ghoshal2024alp}.

\subsection{Neutrinos}
    \label{ssec:forecastnu}

The influence of neutrinos on both the expansion history of the Universe and the growth of cosmic structures enables constraints on the total neutrino mass, $M_\nu$, through galaxy clustering measurements. The most prominent signature is the suppression of the matter power spectrum on scales below the neutrino free-streaming length, which is a cumulative effect arising from the integrated expansion history (see Section~\ref{ssec:neutrinos}). As a Stage-V spectroscopic survey with high tracer number density and large survey volume, MUST is expected to enable precise clustering measurements and place stringent constraints on $M_\nu$, particularly when combined with CMB observations.

\begin{figure}[H]
    \centering
    \includegraphics{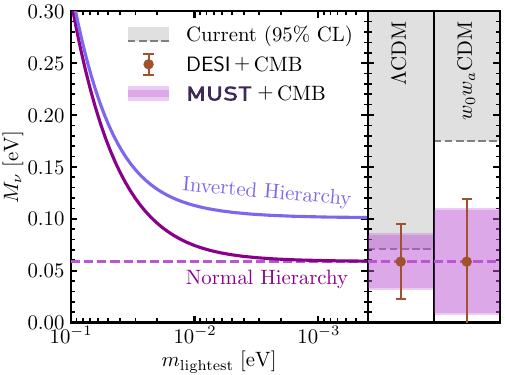}
    \caption{Fisher forecasts for the total neutrino mass, $M_{\nu}$, based on the MUST and DESI galaxy power spectrum with $k_{\rm max} = 0.4\,h\,{\rm Mpc}^{-1}$, combined with existing CMB constraints from Planck PR3 \cite{Planck2020l} and ACT DR6 \cite{ACT2024} data. {\it Left}: $M_\nu$ as a function of the mass of the lightest neutrino species, for both the normal and inverted mass hierarchies. {\it Center}: forecasted $1\,\sigma$ $M_\nu$ uncertainties assuming the fiducial $\Lambda$CDM model. {\it Right}: forecasts assuming the best-fit $w_0w_a$CDM cosmology from \cite{DESI2024c}. The gray shaded region shows the 95\% CL upper bound from the combination of BAO and Planck CMB data \cite{Planck2024}.}
    \label{fig:neutrino_mass}
\end{figure}

Since the total neutrino mass is small, near-future galaxy clustering data alone cannot yield a $>1\,\sigma$ detection of a non-zero $M_\nu$. In such cases, the Gaussian assumption underlying Fisher forecasts breaks down when a physical (positive) prior on $M_\nu$ is imposed. For this reason, we present only forecasts for the uncertainty on $M_\nu$ derived from the combination of galaxy clustering and current CMB observations from Planck and ACT.
Specifically, we perform cosmological inferences using Planck PR3 \cite{Planck2020l} and ACT DR6 \cite{ACT2024} data, and incorporate the resulting covariances into the Fisher matrix derived from our full-shape galaxy clustering forecasts to obtain joint constraints.
We consider both MUST and DESI samples and adopt a fiducial value of $M_\nu = 0.06\,{\rm eV}$, which is approximately the lower bound allowed by the normal hierarchy within the $\Lambda$CDM cosmology.
The forecast results are shown in Figure~\ref{fig:neutrino_mass}. With $k_{\rm max} = 0.4\,h\,{\rm Mpc}^{-1}$, we find that the combination of DESI and CMB data can constrain $M_\nu$ with a percision corresponding to a $\sim1.1\,\sigma$ sensitivity to the mass hierarchy. MUST, when combined with CMB data, further reduces the statistical uncertainty by approximately 25\%, yielding $\sigma(M_\nu) \simeq 0.026\,{\rm eV}$ under both optimistic and conservative assumptions about the available targets.

It is important to note that our forecasts are not directly comparable to current constraints from observational data (e.g., \cite{DESI2024d}). Under the $\Lambda$CDM framework, existing datasets tend to prefer an effective negative neutrino mass, leading to much tighter $M_\nu$ constraints than would be expected from Fisher forecasts \cite{Green2025}. This preference arises because the measured amplitude of matter clustering is higher than predicted by $\Lambda$CDM and can be partially compensated by the inverse suppression effect of an (unphysical) negative neutrino mass. This tension can be significantly alleviated by allowing for a time-evolving dark energy, as discussed in \cite{DESI2024d}.
For this reason, we further perform forecasts assuming the best-fit $w_0w_a$CDM cosmology from \cite{DESI2024c}, as shown in Figure~\ref{fig:neutrino_mass}. In this case, the uncertainties on $M_\nu$ are generally larger due to the increased model freedom. Nonetheless, the joint DESI and CMB constraint is still expected to yield a $\sim1\,\sigma$ detection of a non-zero total neutrino mass, while the combination of MUST and CMB achieves a $\sim$15\% reduction in the $M_\nu$ uncertainty.
In summary, MUST will provide valuable insights into the apparent ``negative neutrino mass'' problem and may offer further cosmological clues about the neutrino mass hierarchy.

\subsection{Warm Dark Matter}
    \label{ssec:forecastwdm}

The Ly$\alpha$ forest is a powerful probe of small-scale structures, offering constraints on dark matter models that suppress small-scale clustering power. When accounting for the effects of inhomogeneous \ion{H}{I} reionization \cite{Montero-Camacho2019b}, the Ly$\alpha$ forest power spectrum can distinguish between different dark matter scenarios even on relatively large scales ($k\lesssim0.6\,h\,{\rm Mpc^{-1}}$) (Zhang et al. in prep.). In this work, we focus on warm dark matter models, whose characteristic suppression of small-scale clustering is representative of a broader class of models with similar cutoff behavior. We forecast the ability of MUST to place competitive constraints on the WDM particle mass by leveraging the reionization imprints on the Ly$\alpha$ forest power spectrum.

Following \cite{Montero-Camacho2019b}, the three-dimensional flux power spectrum is given by
\begin{equation}
\label{eq:plya}
P^{\rm 3D}_{\rm F}(\boldsymbol{k},z,m_{\rm X}) = b^2_{\rm F} (1 + \beta_{\rm F} \mu^2)^2 P_{\rm m} + 2 b_{\rm F} b_\Gamma (1 + \beta_{\rm F} \mu^2) P_{{\rm m}, \psi},
\end{equation}
where $b_{\rm F}$ and $\beta_{\rm F}$ are the flux bias and RSD parameter, respectively. $b_\Gamma$ is the radiation bias \cite{2015JCAP...12..017A}, $\mu \equiv k_\parallel / k$ is the cosine of the angle between the wavevector and the line of sight, and $P_{{\rm m},\psi}$ is the cross-power spectrum between matter and the IGM transparency, which captures the response of the IGM to reionization.  
The second term in Eq.~\eqref{eq:plya} links the small-scale suppression caused by warm dark matter during reionization to a distinct large-scale signature in the Ly$\alpha$ forest, i.e., the impact of inhomogeneous reionization. These reionization imprints are associated with the scale of ionized bubbles, which typically range from a few to several tens of Mpc. On such large scales, the need of nonlinear modeling and the dependence of the bias terms on $m_{\rm X}$ can be reasonably neglected.

\begin{figure}[H]
    \centering
    \includegraphics{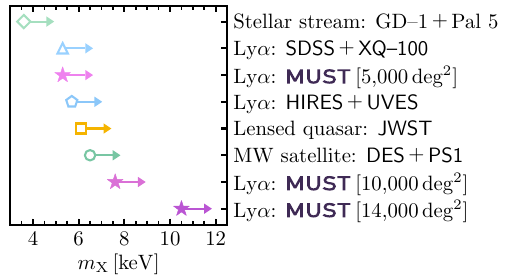}
    \caption{Warm dark matter constraints. Arrows indicate the allowed parameter space (lower bounds) at the 95\% confidence level (CL), which extend towards infinity, corresponding to the CDM limit—the fiducial model for the forecasts. For comparison, we include representative constraints from MW stellar streams \cite{Banik2021}, MW satellite counts \cite{Dekker2022, Nadler2021a}, Lyman-$\alpha$ forest measured from eBOSS $+$ XQ-100 \cite{Palanque2020} and HIRES $+$ UVES \cite{2024PhRvD.109d3511I}, as well as strongly lensed quasars from JWST \cite{Keeley2024}.}
    \label{fig:wdm}
\end{figure}

To forecast constraints on the WDM mass $m_{\rm X}$ from MUST, we include both $m_{\rm X}$ and $\sigma_8$ as free parameters to account for potential degeneracies. The analysis considers 12 redshift bins from 2.0 to 3.8 and 45 $k$ bins spanning 0.09 to 0.67\,$h\,{\rm Mpc}^{-1}$. We adopt an estimated quasar luminosity function for MUST with $r < 23.5$ and evaluate the results with three survey areas: 5,000, 10,000, and 14,000\,deg$^2$. The covariance matrices are computed following \cite{Montero-Camacho2021,2014JCAP...05..023F}, including an aliasing term that accounts for the sparse sampling of quasars.
Figure~\ref{fig:wdm} shows the $m_{\rm X}$ constraints from MUST for different survey areas, along with a comparison to current limits from existing observational data. We adopt a fiducial model corresponding to the CDM limit ($m_{\rm X} \to \infty$). In this case, we constrain the lower bound of the WDM particle mass. An optimistic survey covering 14,000\,deg$^2$ would yield the most stringent lower bound on $m_{\rm X}$ to date, with a 95\% CL limit of 10.5\,keV. Even a conservative 10,000\,deg$^2$ survey would achieve $m_{\rm X} > 7.6$\,keV, exceeding the current strongest Ly$\alpha$ forest-only constraint of $m_{\rm X} > 5.7$\,keV \cite{2024PhRvD.109d3511I}.
We note that further improvements are foreseen with Lyman-$\alpha$ forests from LBGs (e.g., \cite{Herrera-Alcantar2025}), and leave relevant developments to a future work.

\section{Conclusions}
    \label{sec:conclusion}

The MUltiplexed Survey Telescope (MUST) is a 6.5-meter telescope dedicated to multi-object spectroscopic observations.
It features a modularized focal plane with 336 triangular modules that accommodate over 20,000 fibers at the Cassegrain focus.
The spectrographs are designed with three channels to cover a wavelength range of $\sim 3700$--$9600$\,\AA{} at a resolution of $R\sim 2000$--$4000$.
Initial monitoring of observing conditions indicates that the MUST candidate site, Peak A of the Saishiteng Mountain near Lenghu Town, could provide over 2,400 observing hours annually, making MUST at least 10 times more efficient than currently operating spectroscopic surveys.
As a result, MUST is expected to collect over 100 million spectra that are significantly fainter than those in existing spectroscopic databases in the 2030s.

While designed as a flexible spectroscopic platform, the primary mission of MUST is to conduct the first Stage-V galaxy spectroscopic survey\footnote{proposed by the Snowmass Cosmic Frontier report \cite{Chou2022} as a continuation of the stages defined by the Dark Energy Task Force \cite{Albrecht2006DETF}} to address fundamental questions in cosmology and physics.
With its unprecedented photon collecting capability as a spectroscopic telescope, MUST will produce the first 3D map of galaxies spanning from the nearby Universe out to $z \sim 5.5$, which corresponds to $\sim 1$ billion years after the Big Bang.
This large survey volume enables in-depth studies of the dynamic evolution of the dark Universe and signatures of fundamental physics from the primordial era. Thus, the primary scientific objectives of MUST include, but are not limited to, investigating the nature and evolution of dark energy, testing gravity theories, probing inflation physics through primordial non-Gaussianity, and exploring the properties of neutrinos and dark matter.

Different types of targets are required to cover the extensive redshift range with sufficient density to achieve these ambitious scientific goals. For the low-redshift Universe ($z \lesssim 1.6$), MUST will utilize the same galaxy targets and similar selection criteria as DESI, but will focus on fainter samples to avoid redundant observations. Beyond $z \sim 2$, MUST will mainly rely on LBGs and LAEs to map large-scale structures. Although the actual target selection and redshift measurement efficiencies are not yet fully verified, current multi-band imaging and spectroscopic validation data in limited areas suggest that these tracers will be abundant enough for precise cosmological inferences. Meanwhile, QSOs will play a vital complementary role in probing the IGM, which is crucial for studying specific physical effects, such as the mass of dark matter particles.

We have performed forecasts for several key science cases of MUST. The increased galaxy density at low redshift ($z \lesssim 1.5$) and the introduction of new tracers at high redshift ($z \gtrsim 2$) enable percent-level measurements of cosmological distances and structure growth across the entire redshift range of $0 < z \lesssim 5.5$. As a result, constraints on density parameters (including $\Omega_{\rm m}$ and $\Omega_k$), the dark energy equation of state ($w_0$, $w_a$), and the phenomenological modified gravity parameter $\mu_0$ will be substantially tighter than those forecasted for DESI, the current state-of-the-art Stage-IV survey.
This demonstrates the potential of MUST not only to distinguish among different dark energy and gravity models, but also to provide stringent cross-checks of observational systematics when compared with other cosmological probes such as CMB.
With MUST data, constraints on the amplitude of local-type primordial non-Gaussianity ($f_{\rm NL}^{\rm local}$), when combined with current CMB results, are expected to reach the $\sim 1$ level. Even stronger constraints may be achievable with more advanced summary statistics and data analysis methods, offering a powerful test of inflation models.
The combination of MUST and CMB data is further expected to yield a statistical uncertainty of $\sim 0.03\,{\rm eV}$ on the total neutrino mass within the $\Lambda$CDM framework, enabling valuable insights into the absolute neutrino mass and the mass hierarchy.
Additionally, using Lyman-$\alpha$ forest data from quasar spectra, MUST can place a lower limit of $\sim 10.5\,{\rm keV}$ on the warm dark matter particle mass for a $14,000\,{\rm deg}^{2}$ survey area, surpassing the current best constraints by nearly a factor of two.
Together, these forecasts demonstrate that the MUST spectroscopic survey will make a substantial contribution to addressing some of the long-standing fundamental questions in cosmology and particle physics.

\Acknowledgements{
The MUST project is supported by the Ministry of Science and Technology, China (Grant No. 2023YFA1605600) and the Ministry of Education, China.\\ \\
We want to express our sincere gratitude to Mr. Tian-Pei Chen for his generous donation and unwavering trust, which have significantly supported the MUST project. We also thank Mr. Dong-Hua Dong for his donation.\\ \\
The authors thanks Noah Sailer, Anand Raichoor, Christophe Y\`{e}che, Teppei Okumura, Ji Yao for useful discussions.
Cheng Zhao acknowledges the support from the National Natural Science Foundation of China (NSFC) grant No. 12573002.
Song Huang acknowledges the support from the National Natural Science Foundation of China (NSFC) Grant No. 12273015 \& No. 12433003 and the China Crewed Space Program through its Space Application System.
JPK, AR, DF, JY, AV, RG, and SH acknowledge the support from the SNF 200020\_175751 and 200020\_207379 ``Cosmology with 3D Maps of the Universe'' research grant.
Si-Wei Zou acknowledges the support from the National Science Foundation of China (No. 12303011).
Yu Liu acknowledges the support from the National Science Foundation of China (No. 12303005) and the Shuimu Tsinghua Scholar Program (No. 2022SM173).
Pablo Renard acknowledges the support by the Tsinghua Shui Mu Scholarship, funding of the National Key R\&D Program of China (grant no. 2023YFA1605600), and the science research grants from the China Manned Space Project with No. CMS-CSST2021-A05, the Tsinghua University Initiative Scientific Research Program (No. 20223080023), and the National Science Foundation of China (No. 12350410365).
}

\InterestConflict{
    The authors declare that they have no conflict of interest.
}

\bibliography{main}
\bibliographystyle{scpma_short}

\end{multicols}
\end{document}